\documentclass[pre,aps,twocolumn,showpacs]{revtex4}

\usepackage{graphicx}
\usepackage{amsmath}
\usepackage{amsfonts}
\usepackage{amssymb}
\usepackage{epsfig}

\usepackage{epstopdf}
\newcommand{\beqs}{\begin{eqnarray*}}
\newcommand{\eeqs}{\end{eqnarray*}}
\newcommand{\beq}{\begin{eqnarray}}
\newcommand{\eeq}{\end{eqnarray}}

\newcommand{\bit}{\begin{itemize}}
\newcommand{\eit}{\end{itemize}}
\newcommand{\eps}{\varepsilon}
\newcommand{\p}{\partial}

\newcommand{\bu}{\mathbf{u}}

\newcommand{\bt}{\hat{\mathbf{t}}}
\newcommand{\bn}{\hat{\mathbf{n}}}

\begin{document}
\title{Interplay of internal stresses, electric stresses and surface diffusion in polymer films}
\author{Fabien Closa}
\author{Falko Ziebert}
\author{Elie Rapha\"el}
\affiliation{Laboratoire de Physico-Chimie Th\'eorique - UMR CNRS Gulliver 7083, 
ESPCI, 10 rue Vauquelin, F-75231 Paris, France}
\date{\today}
\begin{abstract}
We investigate two destabilization mechanisms for elastic polymer films
and put them into a general framework: first, instabilities due to in-plane stress
and second due to an externally applied electric field normal to the film's free surface.
As shown recently, polymer films are often stressed
due to out-of-equilibrium fabrication processes as e.g.~spin coating. Via an
Asaro-Tiller-Grinfeld mechanism as known from solids, 
the system can decrease its 
energy by undulating its surface by surface diffusion of polymers and thereby relaxing stresses.
On the other hand, application of an electric field is widely used
experimentally to structure thin films: when the electric Maxwell surface stress  
overcomes surface tension and elastic restoring forces, the system undulates
with a wavelength determined by the film thickness.
We develop a theory taking into account both mechanisms simultaneously and
discuss their interplay and the effects of the boundary conditions both at the 
substrate and the free 
surface.
\end{abstract}
\maketitle

\section{Introduction}
\label{Intro}

The stability of polymer thin films is an important research subject
in polymer physics and materials science. On the one hand, the aim may be to obtain
a stable film, as in coatings and lubrification. On the other hand,
soft films are used for microstructuring,
where they are destabilized to yield well-designed patterns
that are used e.g. as a mould for further microfabrication processes. 
In both cases it is crucial to understand the stabilizing and destabilizing
mechanisms that prevail in polymer films, which can be either internal
(Van der Waals forces due to reduced dimensions, internal stresses,
decomposition in mixtures) or external (external stresses, external fields).

In recent studies on spin-coated polymer films it became apparent that
thin films are prone to store residual stresses \cite{Croll1979,Reiter:2005}. 
Such stresses are created due to the fast evaporation
process of the spin-coating process: as evaporation is fast,
the polymer chains do not have the time to reach their equilibrium
configurations and in the final, glassy state the film 
has frozen-in non-equilibrium configurations that give rise to stresses. 
If these stresses are not relaxed, e.g. by ageing or tempering the films,
they influence the film stability as recently shown 
in dewetting experiments and discussed theoretically
\cite{Reiter:2005,Fretigny,Raphael:2006.1,Reiter:2007.2,FZER1}. 
There it has been shown that stresses increase the initial
dewetting velocity and also strongly influence the long time dynamics
of the dewetting films.  
In case the film does not dewet, the stresses may still lead to destabilization
\cite{Raphael:2006.3}, as they should give rise
to an Asaro-Tiller-Grinfeld instability \cite{asaro:72,grinfeld:86,grinfeld:93}. 
This mechanism has been proposed for stressed solids 
in contact with their melt or for solids which evolve via 
surface diffusion. Its origin is the fact that the solid can relax 
stress and lower its energy  
by creating surface undulations. 
For polymer thin films 
the interplay
between residual stresses and other, e.g.~externally applied,
destabilization forces constitutes an interesting question
of importance for all further manipulations of freshly spin-coated films.

In this work we reformulate the energy approach usually used to describe the 
Grinfeld instability in a way that highlights the connection
with other known instabilities in thin films. 
We use the bulk elastostatic equations together with
a time-dependent kinematic boundary condition at the free interface.
A direct coupling term between the height of the polymer film
and the displacement field arises, which has not been discussed before
as it is less relevant in atomic solids. In polymer films, however, this coupling
should be present and important. Moreover this term establishes the connection 
to other elastic instabilities, namely to a buckling-like instability 
under compressive stress and, in the case of an externally applied field, to the elasto-electric instability
investigated by Sharma {\it et al.}~\cite{SharmaPRL,monch2001,Sharmalong,Sharma08}.  
Finally the growth rate of the height of the polymer film
is derived in case of simultaneous action of stress and external field. 
This result is briefly compared to recent experiments
concerning the electrohydrodynamic instability
of very viscous (high molecular weight) spin-coated 
thin polymer films heated above the glass transition \cite{Barbero09,SteinerEPL}.

The work is organized as follows:
First, in section \ref{Grin_elastic} we recall the classical, energy-based 
formulation of the Grinfeld mechanism.
In section \ref{nonlin}, we start from a nonlinear elastic theory, derive
the bulk elastic equations and investigate in section \ref{elastsharma}
the stability under stretch/compression.
In section \ref{grindiffuse} we show that by allowing surface diffusion 
via a kinematic boundary condition for the height of the film, 
the Grinfeld result is regained in a well-defined limit. 
The coupling between height and displacement via the kinematic boundary 
condition can influence the classical Grinfeld instability for intermediate stresses.
In section \ref{grinE} we add the external electric field to our description.
We regain the instability discussed in Sharma {\it et al.}~\cite{SharmaPRL}
in a certain limit. Moreover the full growth rate of the film height is calculated and
its consequences for experiments are briefly discussed. 

\section{Grinfeld instability - classical way of calculation; 
effects of boundary condition at the substate}
\label{Grin_elastic}

To start with we briefly review the classical treatment of the Grinfeld instability
of an elastic medium under uniaxial stress \cite{Nozbook,cantat:98}.
Usually a semi-infinite solid is investigated, but in view of the thin
film geometry we allow for a finite thickness $h_0$ of the film.
The known results for the semi-infinite case can then be obtained
by performing the limit $h_0$ towards infinity.
As the dynamics of this instability is energy-driven
-- the system can lower its energy by creating surface undulations -- 
all the information needed to describe the system is contained in the (free) energy of the system,
which has an elastic part, $E_{{\rm el}}$, and a surface part, $E_{{\rm surf}}$. 

\begin{figure}[t]
\centering
\vspace{.2cm}
\includegraphics[width=0.48\textwidth]{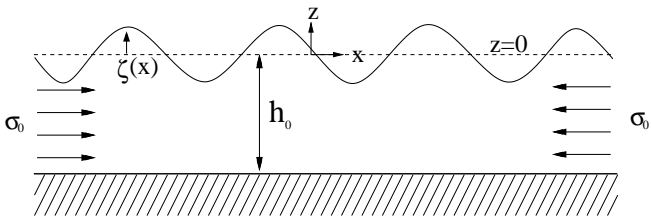}
\caption{\label{fig1} Sketch of the geometry. The thickness of the film
spans from $-h_0$ to $\zeta(x)$ in $z$-direction. There is uniaxial
stress in $x$-direction, which can be either compressive (as shown) or tensile.}
\end{figure}

We chose the coordinate system in such a way that the free surface is at $z=0$,
see the sketch of the geometry in Fig.~\ref{fig1}.
For simplicity we assume a plane strain situation \cite{Maugis}
where the uniaxial prestress $\sigma_0$ is taken along the $x$-axis.
Consequently, we consider an undulation of the surface along $x$ given by
\beq
\zeta(x)=\eps A\cos(kx)\,.
\eeq
We assume either an infinite system or periodic boundary conditions in 
$x$-direction. $k$ is the wave number of the perturbation,
$A$ its amplitude and $\eps$ a small book-keeping parameter used in the following
when dealing with expansions.

The elastic energy of an linearly elastic solid  can be
written via the stress field $\sigma_{\alpha\beta}$ as \cite{landau_el}
\beq
E_{\rm el}=\dfrac{1}{2 E}\int 
\left[ \left(1 + \nu \right) \sigma_{\alpha \beta}^{2} - \nu \sigma_{\alpha \alpha} \sigma_{\beta \beta} \right] dx\,dz\,,
\eeq
with $E$ the elastic or Young's modulus.
Summation convention is implied for indices occurring twice ($\alpha,\beta=1..3$).
Using a plane strain approximation, one gets
\beq\label{Eelplanestr}
E_{\rm el}=\dfrac{1}{2 \bar{E}}\int 
\left[ \left(1 + \bar{\nu} \right) \sigma_{ij}^{2} - \bar{\nu} \sigma_{ii} \sigma_{jj} \right] dx\,dz\,,
\eeq
where now $i,j=1..2$  
($1\leftrightarrow x,2\leftrightarrow z$) and
$\bar{E}=\frac{E}{1-\nu^{2}}$ and $\bar{\nu}=\frac{\nu}{1-\nu}$.
Assuming incompressibility, i.e. a Poisson's ratio of $\nu=1/2$,
one gets
$\bar{E}=\frac{4}{3}E$, $\bar{\nu}=1$. 
We also will use the shear modulus $G$ later on and note the known relations, 
$G=\frac{E}{2(1+\nu)}=\frac{1-\nu^2}{2(1+\nu)}\bar{E}=\bar{E}/4$.
As the system is invariant 
in $y$-direction, $E_{{\rm el}}$ has units of energy per unit length.

The second energy in the problem is the surface energy
\beq\label{Esurf}
E_{{\rm surf}}=\gamma\int\left(\sqrt{1+\zeta'(x)^2}-1\right)dx\,,
\eeq
where $\gamma$ is the surface tension and $E_{{\rm surf}}$ is measured
with respect to the state of a flat surface.

To evaluate the elastic energy, one has to solve the elastostatic problem.
The prestress is 
uniaxial along the $x$-axis and given
by $\sigma^0_{xx}=\sigma_0$, $\sigma^0_{zz}=0$ and $\sigma^0_{xz}=\sigma^0_{zx}=0$.
Note that $\sigma_0<0$ holds for the case of a compressive stress
and $\sigma_0>0$ in case of a tensile stress.
Undulations of the surface will give rise to an additional
relaxational stress $\tilde{\sigma}_{ij}$.
The total stress $\sigma_{ij}=\sigma^0_{ij}+\tilde{\sigma}_{ij}$
has to fulfill the Cauchy equilibrium equation
\beq\label{eqeq}
\nabla_i\sigma_{ij}=0
\eeq
and the compatibility equation
\beq
\nabla^2(\sigma_{xx}+\sigma_{zz})=0\,
\eeq
where $\nabla=(\p_x,\p_z)$. As the prestress $\sigma^0_{ij}$ trivially 
fulfills these equations, we
introduce the Airy stress function $\chi(x,z)$ for the relaxational stress
via the known relations \cite{landau_el,Maugis}
\beq\label{stress_via_chi}
\tilde{\sigma}_{xx}=\frac{\p^2\chi}{\p z^2}\,\,\,,\,
\tilde{\sigma}_{zz}=\frac{\p^2\chi}{\p x^2}\,\,\,,\,
\tilde{\sigma}_{xz}=-\frac{\p^2\chi}{\p x \p z}\,\,\,.\,
\eeq
The equilibrium equation is then automatically fulfilled and the compatibility
reduces to $\nabla^2\nabla^2\chi=0$.
This biharmonic equation has to be solved with the following boundary conditions
(BC).
 
At the free surface $z=\zeta(x)$, the normal-normal component of stress has to
balance the surface tension,
while the shear stress has to vanish.
With $\bn$ and $\bt$ denoting the unit vectors normal and tangential to the surface, respectively,
the BC at the free surface read 
\beq 
\hat{n}_{i}\sigma_{ij}\hat{n}_{j}=
\gamma\,\dfrac{\zeta''(x)}{[1+\zeta'(x)^{2}]^{3/2}}\,\,,\,\,\,\,   
\label{BCgrin}
\hat{t}_{i}\sigma_{ij}\hat{n}_{j}=0 \,,
\eeq
or explicitly 
\beq
(\sigma_0+\tilde{\sigma}_{xx})\zeta'^{2}-2\tilde{\sigma}_{xz}\zeta'+\tilde{\sigma}_{zz}
&=&
\gamma\,\dfrac{\zeta''}{[1+\zeta'^{2}]^{1/2}},\quad\nonumber\\
\label{BCfree}
-\tilde{\sigma}_{xz}\zeta'^{2}+\zeta' \left( \tilde{\sigma}_{zz}-\sigma_0-\tilde{\sigma}_{xx}\right)+\tilde{\sigma}_{xz} 
&=&0\,.
\eeq
Note that all the stresses in Eqs.~(\ref{BCgrin}, \ref{BCfree})
have to be evaluated at the interface, i.e. at $z=\zeta(x)$.

At the bottom surface $z=-h_{0}$, where $h_{0}$ is the
film thickness, 
we impose vanishing normal displacement 
\beq\label{BCbot1}
u_z=0 \quad{\rm at}\quad z=-h_{0}\,,
\eeq
meaning that the film is not allowed to detach from the substrate.
As the second BC, we study two possibilities, depending on the preparation of the system:
First, to study the case of possible slippage at the
lower interface, one prescribes 
\beq\label{BCbot2}
{\rm slip\,\,BC:}\quad\quad\sigma_{xz}=0 \quad{\rm at}\quad z=-h_{0}\,,
\eeq
implying vanishing shear stress at the bottom 
(or equivalently a vanishing force on the lower surface of the film
in $x$-direction, i.e. no traction force).
This condition will be called 'slip BC' in the following. 
A second relevant situation, applying to 
the case where the polymer film is rigidly attached to the lower surface,
will be referred to as 'fixed BC',
\beq\label{BCbotfixed}
{\rm fixed\,\,BC:}\quad\quad u_x=0 \quad{\rm at}\quad z=-h_{0}\,.
\eeq
We will see that these two different BC, slip vs.~fixed, have a qualitative influence on 
the instabilities discussed in the following.

{\bf Slip BC at the bottom:} The solution of the elastostatic problem 
with the slip BC at the bottom, Eqs.~(\ref{BCbot1}, \ref{BCbot2}),
is the Airy stress function Eq.~(\ref{chisol}) given in appendix \ref{det}.
The coefficients occurring therein 
have to be determined by the BC at the free surface: 
one calculates the stresses via
Eqs.~(\ref{stress_via_chi}), evaluates them at the free surface 
$z=\zeta(x)$ and expands in powers of $\epsilon$.
From the BC at the free surface, Eqs.~(\ref{BCfree}),
one then determines the coefficients in the Airy stress function
at order $\mathcal{O}(\epsilon)$, yielding Eqs.~(\ref{asbs}). 

The problem is now solved at linear order in the undulation,
and we can study the corresponding energy of the system. 
The elastic energy will change due to the undulation-induced
relaxational stress $\tilde{\sigma}_{ij}$. This change, 
$\Delta E_{\rm{el}}=E_{\rm{el}}-E^0_{\rm{el}}$, 
explicitly reads 
\beq\label{en_el}
\Delta E_{\rm{el}}&=&\frac{1}{2\bar{E}}\int \bigg[
\left(\sigma_0+\tilde{\sigma}_{xx}\right)^2+\tilde{\sigma}_{zz}^2+4\tilde{\sigma}_{xz}^2 \nonumber\\ 
&&\hspace{1.4cm}-2(\tilde{\sigma}_{xx}  +\sigma_0)\tilde{\sigma}_{zz}-\sigma_0^2\,\bigg] dx\,dz\,.\,\,\,\,\,
\eeq
With the Airy stress function determined, the stress field can be evaluated.
The integrations in Eq.~(\ref{en_el}) have first to be performed 
over the film thickness, $\int_{-h_{0}}^{\zeta(x)} dz$. 
Then one usually averages over $x$, assuming periodic boundary conditions:
by writing $\langle E\rangle$ it is understood that one has
averaged like $\frac{k}{2\pi}\int_{0}^{2\pi/k}\,dx$.
Note that due to this averaging the contribution in $\mathcal{O}(\epsilon)$
vanishes.
To leading order $\mathcal{O}(\epsilon^{2})$ one calculates
\beq
\langle\Delta E_{\rm{el},s}\rangle\hspace{-1mm}&=&\hspace{-1mm}
-\frac{A^2}{2}\frac{k\left[\sigma_{0}^{2}+k^{2}\gamma^{2}
+\left(\sigma_{0}^{2}-k^{2}\gamma^{2} \right)\cosh\left(2 h_0 k \right)\right]}
{\bar{E}\left(2h_0k+\sinh\left(2 h_0 k\right)\right)}\,.\,\,\,\,\nonumber\\
\eeq
The surface energy is directly calculated from 
Eq.~(\ref{Esurf}) and yields in order $\mathcal{O}(\epsilon^{2})$
\beq
\langle E_{{\rm surf}}\rangle=\gamma\langle\frac{1}{2}\zeta'(x)^2\rangle=\frac{\gamma}{4} A^2 k^2\,.\,\,\,
\eeq
Note that the averaged quantities, $\langle\Delta E\rangle$, have 
units of energy per unit area. 

Let us now briefly discuss the obtained result.
To regain the classical limit of an semi-infinite elastic half space 
one performs the limit $h_{0}\rightarrow\infty$.
The change in total energy, 
$\langle \Delta E_{{\rm tot}}\rangle=\langle \Delta E_{\rm{el}}+E_{{\rm surf}}\rangle$,
then reduces to 
\beq\label{DEhinf}
\langle \Delta E_{{\rm tot},s}\rangle=-\frac{A^2}{4}\left(\frac{2\sigma_0^2}{\bar{E}}k-\gamma k^2\right)\,.\,\,\,
\eeq
As becomes apparent, the 
prestress 
leads to an {\it decrease} of the energy.
The stress enters quadratically, thus both compressive and tensile
stress trigger the instability which makes it
different from buckling instabilities \cite{landau_el}, see also section \ref{elastsharma}. 
The surface tension acts against the instability and stabilizes high
wave numbers, see the second term in Eq.~(\ref{DEhinf}). 

As we are predominantly interested in polymer thin films, let us perform
the opposite limit, $h_0 k\ll1$. 
This amounts to saying that the modulation wavelength is large compared
to the film thickness.
In this 'thin film' limit, the change in total energy reads
\beq\label{DEhthin}
\langle \Delta E_{{\rm tot},s}\rangle
=-\frac{A^2}{4}\left(\frac{\sigma_0^2}{\bar{E}h_0}\left[1+\frac{2}{3}(h_0 k)^2\right]-\gamma k^2\right).\,\,\,\,\,\,
\eeq
Again, the prestress is destabilizing, independent of its sign. 
However, there is no wave number dependence of the destabilizing term to leading order. 
The same result was recently obtained in the framework of 
a lubrication approximation \cite{Raphael:2006.3}.

{\bf Fixed BC at the bottom:} In this case
the Airy stress function given by Eqs.~(\ref{chisolf}, \ref{afbf})
has to be used.
For the change in the elastic energy this results in
\beq
&&\hspace{-5mm}\langle\Delta E_{\rm{el},f}\rangle\nonumber\\
&=&
-\frac{A^2}{2} \frac{k\left[ (\sigma_0^2-k^2 \gamma^{2}) \sinh\left(2 h_0 k\right)
+ 2 h_0 k(\sigma_0^2+k^2 \gamma^{2}) \right]}
{\bar{E}  \left(2 h_0^2 k^2+\cosh (2 h_0 k)+1\right)}\,.\,\,\,\,\nonumber\\
\eeq
To cross-check, in the classical limit of a semi-infinite elastic half space 
one again recovers Eq.~(\ref{DEhinf}). This is expected
as for a half space the BC at the bottom 
should not be important.
In contrast, in the thin film limit $h_0 k\ll1$ one gets
\beq\label{DEhthinfix}
\langle \Delta E_{{\rm tot},f}\rangle
=-\frac{A^2}{4} \left(\frac{4\sigma_0^2 }{\bar{E}h_0} (h_0 k)^2-\gamma k^2\right).\,\,\,\,\,\,
\eeq
Note that the prestress still lowers the energy, but now has the same wave number
dependence as the contribution from surface tension.
Thus only above a threshold,
\beq\label{thresh_fixed}
\sigma_0>\sqrt{\frac{\gamma\bar E}{4h_0}}\,,
\eeq
the prestress can destabilize the system.

We have seen that the total energy of the system can be lowered
by surface undulations in all the cases discussed above.
To establish these favorable undulations, it needs a mechanism that allows rearrangements to occur.
In the classical case of a solid in contact with its vapor, this is achieved by 
melting-crystallization processes at the surface.
This results in a velocity of the boundary $v_{MC}=\Gamma\Delta E$
\cite{grinfeld:86}, where $\Gamma$ is a mobility.

A second possible mechanism -- on which we would like to focus here
in view of polymers -- is {\it surface diffusion}.
If atoms or vacancies (in case of a solid) or polymer chains
(in case of polymer films) feel an inhomogeneous chemical potential
at the surface, they will diffuse. As a result, the boundary will move with
a velocity $v_{D}=-Mk^2\Delta E$,
with a mobility coefficient $M$ \cite{asaro:72}.
Note the second order spatial derivative stemming from the diffusion
process and reflected in the $k^2$-dependence of $v_{D}$. 

Much more is known about the Asaro-Tiller-Grinfeld-instability, for which 
we refer to the literature. 
For the nonlinear evolution beyond the instability, see 
\cite{nozieres:93,Kohlert:03} for analytical work and \cite{Kassner:01}  
for phase-field modeling. 
Concerning experiments, very clean realizations of this instability have been observed 
in Helium crystals \cite{torii:92} and single crystal polymer films \cite{Berrehar92}.

\section{Stretched elastic solid: nonlinear bulk formulation}
\label{nonlin}

To properly describe finite stresses in a thin polymer film, one
has to use a nonlinear elasticity formulation. 
Let us assume that the film was originally in a stress-free state, described by coordinates
$\mathbf{X}=X_{i}\mathbf{e_{i}}$. Then we stretch (or compress) the film, 
for simplicity uniaxially in the $x$-direction by a factor $\lambda>1$ ($\lambda<1$). 
This state will be described by  coordinates $\mathbf{x}=x_{i}\mathbf{e_{i}}$ and considered
as the {\it base state}. This state will be under uniaxial stress $\sigma_{xx}^0$, see below.
Finally the film is brought in close contact with the substrate (either still permitting for slip,
or perfectly fixed to it, see the two BCs discussed in the last section) and we let it evolve. 
This current state will be described
by coordinates $\mathbf{\tilde{x}}=\tilde{x}_{i}\mathbf{e_{i}}$.
Note that we discuss here only the simple situation where the film is 
attached {\it after} the stretch. The situation where the polymers
attach to the substrate while the film is stretched (which probably better corresponds to
the situation during spin-coating) is more involved as the uniaxiality
is broken due to the presence of the substrate, cf. Ref.~\cite{ken89}.

The total deformation gradient from $\mathbf{X}$ to $\mathbf{\tilde{x}}$ 
reads 
\beq
\mathbf{F}=\frac{\p\mathbf{\tilde{x}}}{\p\mathbf{X}}
=\frac{\p\mathbf{\tilde{x}}}{\p\mathbf{x}}\cdot\frac{\p\mathbf{x}}{\p\mathbf{X}}
=:\mathbf{F_{2}}\cdot\mathbf{F_{1}}\,.
\eeq
Here 
\beq
\mathbf{F_{1}}= \mathrm{Diag}(\lambda, \lambda^{-1/2},\lambda^{-1/2})
\eeq 
describes the stretching (compression) of the film  by a factor
$\lambda>1$ ($\lambda<1$). Note that this step must be described in the nonlinear regime,
as stresses are finite.  
The second tensor (with $\mathbf{I}$ the identity),
\beq
\mathbf{F_{2}}=\mathbf{I}+\nabla\mathbf{u}\,,
\eeq
introduces the usual linear displacement gradient tensor $\nabla \mathbf{u}=(\p_ju_i)_{ij}$ in the current state 
with respect to the stretched state. 
As we are only interested in the stability of the base state,
here a linearized theory is enough for our purposes.
As usual we denote with $\mathbf{B}=\mathbf{F} \cdotp \mathbf{F}^{\mathrm{T}}$ 
and  $\mathbf{C}=\mathbf{F}^{\mathrm{T}} \cdotp \mathbf{F} $
the left and right Cauchy-Green tensors.
As $\mathbf{B}$ is in Eulerian frame we adopt it for the stresses.
$\mathbf{C}$ is in Lagrangian frame and is more convenient for the energy definition.
Using a Neo-Hookean elastic solid \cite{Macosko} , the Cauchy stress tensor is defined as
\beq\label{Cauchy}
\sigma = G \mathbf{B}-P \mathbf{I}\,.
\eeq 
It describes the stress after a deformation in the current configuration. 
$P$ is a Lagrangian multiplier (an  effective pressure having
units of $[{\rm Pa}]$)
that ensures the incompressibility condition.
In the base state, 
from Eq.~(\ref{Cauchy}) one directly gets $\sigma_{\alpha\beta}^0=0$ except for 
\beq\label{rel_sig_lambda}
\sigma_{xx}^0=G\left(\lambda^2-\lambda^{-1}\right)\,.
\eeq
This establishes a connection between the stretch factor $\lambda$ and the prestress $\sigma_0$.

Now, let us consider again a plane deformation with respect to the prestretched base state. 
Evaluating the Cauchy stress tensor
in linear order in the displacement gradient, imposing plane strain and using incompressibility,
one arrives at the bulk equations 
\beq
G \left( \lambda^{2} \partial_{x}^2u_{x} + \lambda^{-1} \partial_{z}^2u_{x} \right) - \partial_{x}P &=& 0, \nonumber\\
G \left( \lambda^{2} \partial_{x}^2u_{z} + \lambda^{-1} \partial_{z}^2u_{z} \right) - \partial_{z}P &=& 0.
\eeq
Note the asymmetry introduced by $\lambda\neq1$, i.e. the prestretch. 
All quantities can be expressed either in the base state $\mathbf{x}$ or in the current state 
$\mathbf{\tilde{x}}$ - as deformations $\mathbf{u}$ are small, they amount to the same expressions.
For $\lambda=1$ one regains the classical
elastostatic equation for an incompressible solid,
$G \nabla^2 \bu+\nabla P=0\,,$
where $\bu$ is the displacement field.

The elastic energy density, $\rho_{\rm el}$, for the Neo-Hookean elastic solid reads
\beq
\label{Neo-Hookean}
\rho_{\rm el}= \dfrac{G}{2} \left( \mathrm{Tr}(\mathbf{C}) - 3 \right)\,.
\eeq
Here we did not include the pressure as a Langrangian multiplier (giving rise
to a term $+P(\det(\mathbf{B})-1)$), as incompressibility is imposed when solving the bulk equations,
see the next section.
Note that for plane strain and small deformations 
one regains Eq.~(\ref{Eelplanestr}) to second order in displacement gradients, i.e.
$\rho_{\rm el}=\dfrac{1}{2 \bar{E}}\left[ \left(1 + \bar{\nu} \right) \sigma_{ij}^{2} - \bar{\nu} \sigma_{ii} \sigma_{jj} \right].$

Now we have established the equations for a nonlinear prestretch and a subsequent linear theory.
Note, however, that the Neo-Hookean model should not be used for $\lambda$-values too far from $\lambda=1$.
Otherwise effects of e.g.~the crosslink length must be taken into account
and one should use more realistic models like the Mooney-Rivlin solid \cite{Macosko}.
We will now investigate the stability of the prestressed base state 
with respect to surface undulations:
(i) for the purely elastic case, (ii) in the presence of surface diffusion,
making a connection with the classical Grinfeld instability, (iii)
in the presence of an electric field normal to the free surface, regaining and generalizing results obtained
previously \cite{SharmaPRL,HeEPL} and finally (iv) with both surface diffusion and applied electric field. 

\section{Stretched elastic solid: solutions for surface modulations}
\label{elastsharma}

In this and the following section we solve the elastic bulk equations 
and show how surface diffusion can be incorporated within this approach 
in a generic way to regain and generalize the Grinfeld result.
We use the same boundary conditions as introduced in section \ref{Grin_elastic}, i.e.
Eq.~(\ref{BCfree}) for the free surface and either 
the slip BC or 
the fixed BC at the bottom.

The stability of the base state can be studied by the ansatz 
\beq
u_x(x,z,t) &=& u_x(z) e^{ikx+st} \nonumber\\
u_z(x,z,t) &=& u_z(z) e^{ikx+st} \nonumber\\
P(x,z,t) &=& p(z) e^{ikx+st}
\eeq
where the amplitudes 
are small perturbations of order $\mathcal{O}(\eps)$ in height perturbations, see Eq.~(\ref{heightperturb}) below.
Note that we allowed for a temporal dependence which will be used only in the following sections.
Using incompressibility, $iku_x+\partial_z u_z = 0 $, one obtains 
a single decoupled equation for $u_z$ given by
\beq\label{uzeq}
k^4 \lambda ^3 u_z-k^2 \left(\lambda ^3+1\right) u_z''+u_z^{(4)}=0\,.
\eeq
With this equation solved, one easily obtains $u_x$ from incompressibility 
and the pressure from
$p(z) = G  \left(\frac{u_{z}'''(z)}{\lambda  k^2}-\lambda ^2 u_{z}'(z)\right)$.
The general solution
of Eq.~(\ref{uzeq}) reads (for $\lambda\neq1$ \footnote{Note that the case $\lambda=1$ is singluar as it yields 
only one wavenumber and additional solutions like $z\sinh(z)$.})
\beq\label{gensol_uz}
u_z(z)&=&\epsilon\sum_{i=1,2}\big\{A_i\cosh \left[ k_i \left( z + h_0 \right) \right]\nonumber\\
&&\quad\quad\quad+B_i\sinh \left[ k_i \left( z + h_0 \right) \right]\big\}\,,
\eeq
with 
\beq
k_1=k\,,\,\,\,{\rm and}\,\,k_2=l=k\lambda^{3/2}\,.
\eeq
Imposing the BCs at the substrate yields
\beq
u_z(-h_0) &=& 0 \Leftrightarrow A_2 = -A_1 \nonumber\\
\sigma_{xz}(-h_0) &=& 0 \Leftrightarrow A_2 = 0 ,\hspace{1cm}\quad({\rm slip\,\,BC})\nonumber\\
u_x(-h_0) &=& 0 \Leftrightarrow B_2 = - \dfrac{k}{l} B_1,\quad({\rm fixed\,\,BC})\,.\nonumber
\eeq
As before, we parameterize the  upper free interface of the thin polymer film by a harmonic function 
with small amplitude of order $\mathcal{O}(\eps)$
\beq\label{heightperturb}
z= h(x,t) = \epsilon h e^{ikx+st}. 
\eeq
The normal vector of this surface reads $\hat{\mathbf{n}} = (-ik h(x,t), 0 , 1)$ at first order.
Thus at the free interface, cf. Eqs.~(\ref{BCgrin}), the BCs read
$\sigma_{ij}  \hat{n}_{j} = - \gamma k^2 h(x,t) \hat{n}_{i}$.
They fix the remaining unknown coefficients and
one obtains $A_{1,s}=A_{2,s}=0$ and
\beq
B_{1,s} &=&  - h \dfrac{2 k \left(G \left(l^2-k^2\right) \cosh (h_0 l)+k^2 l \gamma  \lambda  \sinh (h_0 l)\right)}{G
   \left(k^2+l^2\right)  g^+(k,l)}\,, \nonumber\\
B_{2,s} &=& h \dfrac{k^3 \gamma  \lambda  \sinh (h_0 k)+G  \left(l^2-k^2\right) \cosh (h_0 k)}{ l G  \, g^+(k,l) }\,
\eeq
in case of the slip BC at the bottom;
for the fixed BC 
\beq
A_{1,f} &=& h \dfrac{G  \left(l^2-k^2 \right) v(k,l,k) + k^3 \gamma \lambda w(k,l) }
{G \left(k^4+6 k^2 l^2+l^4 -\left(k^2+l^2\right) \, f(k,l) \right)},\,\,\,\nonumber\\
B_{1,f} &=& - h  \dfrac{G \left(l^2-k^2 \right) w(k,l) + \gamma \lambda k^3 v(k,l,l)}
{G  \left(k^4+6 k^2 l^2+l^4 -\left(k^2+l^2\right) \, f(k,l) \right)}\quad\,\,\,\,
\eeq
and $A_{2,f}=-A_{1,f}$, $B_{2,f} = - \dfrac{k}{l} B_{1,f}$.  
We introduced the following abbreviations 
\beq
g^{\pm}(k,l) &=& \pm\sinh \left[ \left(l + k \right) h_0 \right] \left(l -k \right) \nonumber\\
&&+ \sinh \left[ \left(l  - k \right) h_0 \right] \left(l + k \right)\,, \nonumber \\
f(k,l) &=& \cosh \left[ \left(l + k \right) h_0 \right] \left(l -k \right)^2 \nonumber\\
&&+ \cosh \left[ \left(l  - k \right) h_0 \right] \left(l + k \right)^2\,, \nonumber \\
v(k,l,m) &=& \left( l^2 + k^2 \right) \cosh \left[ k h_0 \right] - 2 m^2 \cosh \left[ l h_0 \right]\,,\nonumber \\
w(k,l) &=& \left( l^2 + k^2 \right) \sinh \left[ k h_0 \right] - 2 k l  \sinh \left[ l h_0 \right]\,.
\eeq

With the general solution obtained, we can now investigate whether the base state is stable or unstable. 
According to Eq.~(\ref{heightperturb}), solutions with non-zero wavenumber, if they exist, correspond to
surface undulations.
The condition for nontrivial solutions to exist can be written
as 
\beq\label{condntsol}
h(x,t)=u_z(x,z=0,t)
\eeq 
or $\epsilon h=u_z(0)$.
Namely, for consistency the displacement at the surface
must equal the height perturbation. An alternative formulation would have been
to write down the system of BCs as a $4\times4$-matrix equation and looking for nontrivial
solutions via the zeros of the determinant.
With $u_z$ known, Eq.~(\ref{condntsol}) can be written as 
\beq
hZ(k)=0
\eeq
with a function of wave number $Z(k)$. If one finds wave numbers $k^*$ with $Z(k^*)=0$, 
periodic solutions exist; otherwise
$hZ(k)=0$ implies $h=0$ and the film stays flat.
Explicitly, for the two considered BCs one gets 
\beq
\label{F_slip}
Z_{s}(k) &=& 4 k^3 l G \cosh\left(l h_0 \right) \sinh\left(k h_0 \right)\nonumber \\ 
& & -\sinh\left(l h_0 \right) \left(l^2 + k^2 \right)^2 G \cosh\left(k h_0 \right) \nonumber
 \\ & & -\sinh\left(l h_0 \right) k^3 \gamma \lambda \left(l^2 - k^2 \right) \sinh\left(k h_0 \right), \\
\label{F_fix}
Z_{f}(k) &=& 4 G k^2 l \left( l^2+k^2 \right)+\frac{G}{2}(r\left(-l,k \right) - r\left(l,k \right))  \nonumber \\
& & + \frac{1}{2} k^3 \gamma \lambda \left( l^2-k^2 \right) g^{-}\left(k,l \right)\,, 
\eeq
with 
$r\left(k,l \right) =(k+l)^2 (k^3 + 3 k l^2 - k^2 l + l^3)\cosh\left[h_0 \left(k - l\right) \right]$. 

For both BCs, nontrivial solutions do not exist under tension, $\lambda>1$, as one would expect.
Buckling occurs under compression, but only for non-physical values,
namely for  $\lambda<\lambda_c\simeq0.03$
for a typical surface tension of $\gamma = 0.5 h_{0} \bar{E}$. 
For such high compressions, the Neo-Hookean law is no longer a good description.
Moreover, the assumption that the film stayed flat in the first step (from $\mathbf{X}$ to $\mathbf{x}$,
i.e. before attaching to the substrate) is not valid anymore - the film would have buckled long before. 
Indeed the threshold for buckling for two free surfaces should be lower
than for the BC that the film does not detach from the substrate surface, Eq.~(\ref{BCbot1}).
Thus we can conclude that the film stays flat for all reasonable values of $\lambda$, $G$ and $\gamma$.
Note, however, that films can be unstable if they are swollen
{\it in the presence} of the substrate, cf.~Refs.~\cite{TanakaNat87,ken89}.

\section{Adding surface diffusion - the Grinfeld instability again and corrections}
\label{grindiffuse}

In the last section we investigated the stability of the base state
with respect to in-plane stresses and found that the purely elastic system is stable.  
Here we add the effects of diffusion of polymer chains
close to the film surface due to stress relaxation-induced changes in the chemical potential. 
As a consequence the system
can produce undulations by {\it diffusive transport} of material,  
in addition to possible elastic displacements. We show that 
one regains the Grinfeld instability in a well-defined limit. 
The overall result is more general as it comprises corrections to the Grinfeld mechanism,
see below.

If we allow for surface diffusion, Eq.~(\ref{condntsol}) has to be modified in order
to allow for this dynamics. 
For the height modulation $h=h(x,t)$ one can write
\beq\label{surdiffus}
\partial_t h&=&\partial_t u_{z|z=0}-(\partial_t u_{x|z=0})(\partial_x h)\nonumber\\
&&+M\partial_x^2\left(\delta \mu_{|z=0}\right)\,.
\eeq
The first two terms on the r.h.s.~stem from the standard kinematic BC
at a free surface, usually written as $\p_t h=v_z-v_x\p_x h$ with $h$ the height of the surface
and  $(v_x,v_z)=\p_t(u_x,u_z)$ the fluid velocity \cite{Bankoff:1997}. 
The second term is purely nonlinear and can be 
neglected in the following linear analysis.

The last term on the r.h.s.~represents the surface diffusion (note that in three dimensions
$\p_x^2$ has to be replaced by the surface Laplacian \cite{Spencer91}). 
It will smoothen gradients in the chemical potential, which is given by 
\beq\label{muwithgam}
\delta\mu=\delta E_{\rm el}-\gamma\kappa\,.
\eeq
$\kappa$ is the mean curvature of the surface, given at $\mathcal{O}(\epsilon)$ by
$\kappa=\p_x^2 h$. $\delta E_{\rm el}$ is the change in elastic energy density due to the surface undulation,
compared to the flat surface.
The coefficient $M$ is a mobility \cite{Mullins,Saul} and explicitly reads 
$M=\frac{D n_s V^2}{k_B T}$,
where $k_B T$ is the thermal energy, $V$ is a microscopic volume 
(of the polymer chain in our case),
$D$ is the surface diffusion coefficient 
and $n_s$ is the surface density of diffusing objects.
Note that in the view of recent experiments on spin-cast polymer melts,
we here allow for a finite chain mobility (at least close to the free surface), 
although we assumed a purely elastic behavior of the film. A generalization
of our approach to the more adequate viscoelastic case will be the subject of a forthcoming study.

Eq.~(\ref{surdiffus}) for the dynamics of the surface undulation
is further motivated in appendix \ref{motkinBC}.
The terms 
arising naturally from the kinematic BC
are commonly not included in the treatment of the
Grinfeld instability, as in the usual context one concentrates
on the diffusive transport of atoms or vacancies. 
Taking the coupling to the displacement into account -- if extended objects like polymers
are diffusing -- corrections to the 'classical' Grinfeld behavior arise:
the time derivative in $\partial_t u_{z|z=0}$ leads to a renormalization of the growth rate $s(k)$
of the height perturbations $h(x,t) = \epsilon h e^{ikx+st}$.
In view of this, in the following we will sometimes compare the 'classical' Grinfeld and the 'kinematic' case.

In the previous section we have already calculated the general solution for the
displacements. Thus the stress tensor is also known and
using Eq.~(\ref{Neo-Hookean}) one gets
the  changes in the elastic energy $\delta E_{\rm el} = \rho_{\rm el} - \rho_{\rm el}^{0}$
with respect to the base state
\beq
\delta E_{el,s} 
&=& -\frac{\epsilon  G  \left(l^2-k^2\right) \cos (k x)}{k^2 \lambda } \nonumber \\
& &\hspace{-1cm}\cdot 
\left(B_{1,s} k \cosh \left[k\left(h_0+z\right)\right]
+B_{2,s} l \cosh \left[l \left(h_0+z\right)\right]\right),\qquad\\
\delta E_{el,f} &=&- \frac{\epsilon  G  \left(l^2-k^2\right) \cos (k x)}{k^2 \lambda } \nonumber \\
& &\hspace{-1cm}\cdot\,\Big(
A_{1,f} \left( k \sinh \left[k (h_0+z)\right]-l \sinh\left[l (h_0+z)\right] \right)\nonumber \\
& &\hspace{-0.8cm}+ B_{1,f} k \left(\cosh \left[k (h_0+z)\right]-\cosh\left[l (h_0+z)\right]\right) \Big).\,
\eeq
For the surface energy, as before Eq.~(\ref{Esurf}) yields
$E_{\rm surf}=\epsilon \cos\left(k x \right) e^{st}\gamma k^2 h$.
Now we can proceed in two ways: 

{\bf Classical calculation, nonlinear case:}
First we can use the classical Grinfeld argument,
i.e.~we integrate from $-h_0$ to $h(x)$ over the film thickness 
and average over the assumed periodic $x$-direction to obtain  
$\Delta E_{\rm el}=\langle\int \delta E_{\rm el} \, \mathrm{d}z\rangle_{x}$.
Upon averaging the linear order in $\epsilon$ vanishes. At $\mathcal{O}(\epsilon^2)$, one gets
to leading order in $k$ 
\beq
\delta E_{el,s} &=& -\frac{h^2 \epsilon ^2 \left(\lambda ^3-1\right)^2 \left(3 \lambda ^3+1\right) G }{16 h_0
   \lambda  \left(\lambda ^3+1\right)^2} + \mathcal{O}(k^{2})\,,\,\,\, \\ 
\delta E_{el,f} &=& -\frac{h^2 k^2 \epsilon ^2 h_0 \left(\lambda ^3-1\right)^2 G}{4
   \lambda } + \mathcal{O}(k^{4})\,. 
\eeq
Let us compare to the result obtained in section \ref{Grin_elastic}.
In the limit $\lambda=1\pm\delta$ with $\delta\ll 1$ and using $\delta=\sigma_0/(3G)$ as
implied by Eq.~(\ref{rel_sig_lambda}) in this limit, one gets including the surface energy
\beq\label{NLGslip}
\delta E_{{\rm tot},s} &=& -\dfrac{h^2}{4}\left(\dfrac{\sigma_0^2}{\bar{E} h_0} \left[1 + \left(h_0 k \right)^2\right]
-\gamma k^2\right), \\
\delta E_{{\rm tot},f} &=&  -\dfrac{h^2}{4}\left(\dfrac{4 \sigma_0^2}{\bar{E} h_0}\left(h_0 k \right)^2-\gamma k^2\right).
\eeq
Note that in leading order this is exactly Eqs.~(\ref{DEhthin}, \ref{DEhthinfix}).
The correction $\left(h_0 k \right)^2\ll 1$ in Eq.~(\ref{NLGslip}) 
has a slightly different prefactor as in Eq.~(\ref{DEhthin}),
which is due to the fact that the fully linear calculation from section \ref{Grin_elastic}
is only correct for {\it infinitesimal} stresses.

{\bf Consistent calculation at order $\mathcal{O}(\epsilon)$:}
The use of an averaging in the Grinfeld calculation seems not necessary to us.
We will thus determine the growth rate of surface undulations by using
\beq
\partial_t h&=&\partial_t u_{z|z=0}+M\partial_x^2\left(\delta \mu_{|z=0}\right)\,.
\eeq
The l.h.s. and the first term on the r.h.s. are of first order in $\epsilon$.
Thus it is sufficient to determine the change of the chemical potential
at this order, i.e. evaluating $\delta\mu$ at the surface.
The full growth rates obtained by this equation are given 
by Eqs.~(\ref{fullss})-(\ref{fullsfkin}) 
in appendix \ref{det}
for the slip and the fixed BC, respectively. 
In the thin film limit $h_0 k\ll 1$, one obtains 
\beq\label{ssk}
s_s (k) &\simeq& M k^2 \dfrac{G (\lambda^3-1)^2}{2 h_0 \lambda (1+\lambda^3)} \nonumber\\
&&\hspace{-1cm}+ M k^4 \bigg[ \dfrac{G h_0 (\lambda^3-1)^2}{6 \lambda}
- \gamma \left( 1 + \dfrac{\lambda^3 - 1}{2 \left( 1 +\lambda^3 \right)} \right)\bigg]\,\,\,
\eeq
for the slip BC at the bottom and neglecting the $\partial_t u_{z|z=0}$ term at the free surface. 
Including the kinematic term yields
\beq\label{sskkin}
s_{s,kin} (k) &\simeq& M k^2 \dfrac{G (\lambda^3-1)^2}{h_0 \lambda (3+\lambda^3)} \nonumber \\ 
& &\hspace{-1.7cm} + M k^4 \bigg[ \dfrac{G h_0 (4+3\lambda^3+\lambda^6) (\lambda^3-1)^2}{3 \lambda (3+\lambda^3)^2}  
- 4 \gamma\dfrac{ (1+\lambda^3)^2 }{(3+\lambda^3)^2}\bigg].\nonumber\\
\eeq
For the fixed BC we get in both cases
\beq\label{ssf}
s_{f} (k) &\simeq& M k^4 \left[\dfrac{G h_0}{\lambda} (\lambda^3-1)^2 -\gamma\right]\,.
\eeq

Let us first discuss the limit of small stresses asgain, $\lambda=1\pm\delta$ with $\delta=\sigma_0/(3G)\ll 1$.
Both Eqs.~(\ref{ssk}, \ref{sskkin}) yield at leading order in the stress
$s(k)\simeq M k^2 \frac{\sigma_0^2}{\bar E h_0}$.
Except for a factor of $4$, at leading order this is exactly Eq.~(\ref{DEhthin}).
The same is true for the fixed BC and Eq.~(\ref{DEhthinfix}). 
Hence in the low-stress and low-wave number limit, our results obtained for the dynamic BC at the free surface
are identical to those obtained by the energy-based calculation in section \ref{Grin_elastic}
in the following sense: $s(k)$ are growth rates as calculated from a dynamical equation 
for the surface undulation.
When comparing to the Grinfeld calculation, there too one 
has to impose a diffusion dynamics driven by the decrease in energy.
One can write 
$\p_t A=-M k^2 \frac{\p}{\p A}\langle \Delta E_{{\rm tot}}\rangle$,
with $E_{{\rm tot}}\propto A^2$. 
The variational derivative with respect to $A$ 
yields a factor of 2.
Taking into account that in the energy approach one
has averaged over $\langle \cos^2(kx)\rangle$ yields another factor of 2,
which explains the differing prefactors.
However, one should note that using the spatial averaging process 
implies a calculation order $\mathcal{O}(\eps^2)$, 
while our method is $\mathcal{O}(\eps)$.

\begin{figure}[t]
\centering
\vspace{.2cm}
\includegraphics[width=0.48\textwidth]{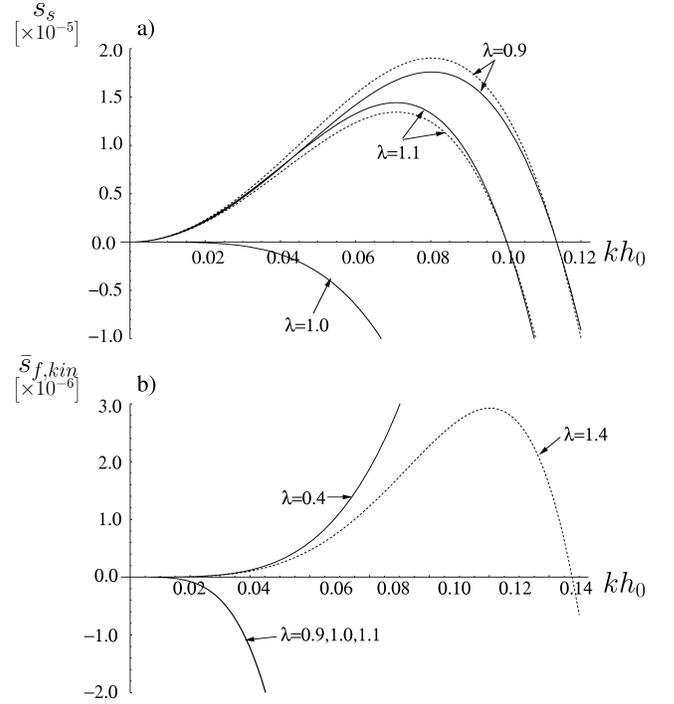}
\caption{\label{ssfig} Growth rates as a function of reduced wave number 
for different stretch factors $\lambda$, corresponding to different prestress. 
Panel a) displays the case of the slip BC at the bottom. Without stress, $\lambda=1$, the system is stable.
Any compression ($\lambda<1$) or stretch ($\lambda>1$) lead to an instability, but with differing rates. 
$\lambda=1.1$ and $\lambda = 0.9$ correspond both to a prestress $|\sigma_0|/\bar{E}\simeq 0.075$.
Solid lines have been obtained with the kinematic BC at the free surface, pointed lines just with
the surface diffusion.
Panel b) shows the case of the fixed BC at the bottom. 
An instability only occurs beyond critical $\lambda$-values.
Parameters: $\gamma = 0.5 h_{0} \bar{E}$.
}
\end{figure}

\begin{figure}[t]
\centering
\vspace{.2cm}
\includegraphics[width=0.48\textwidth]{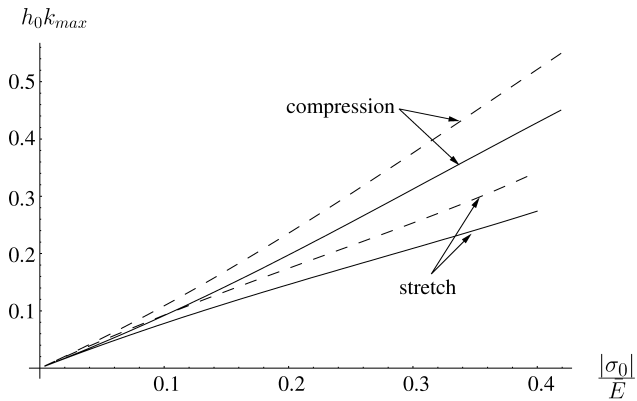}
\caption{\label{kmax} The fastest growing wavenumber $k_{max}$ as given by the maximum
of the growth rate $\bar{s}_{s,kin}$ shown in Fig.~\ref{ssfig}a) as a function of rescaled prestress 
$|\sigma_0|/\bar{E}$. Solid lines are obtained for $\gamma=0.7h_0\bar E$, dashed lines for
$\gamma=0.5h_0\bar E$. 
} 
\end{figure}

Let us now discuss the effect of finite stretches and of the kinematic contribution.
At leading order in the stress, both Eqs.~(\ref{ssk}, \ref{sskkin}) reduce 
to the Grinfeld result. 
However, in next order in the stress Eq.~(\ref{ssk}) yields 
$-\frac{2}{3} M k^2 \frac{\sigma_0^3}{{\bar E}^2 h_0}$,
while the kinematic version yields
$+\frac{1}{3} M k^2 \frac{\sigma_0^3}{{\bar E}^2 h_0}$.
First, this shows that the symmetry with respect to the sign of the stress,
i.e. whether it is due to stretch ($\sigma_0>0$) or compression ($\sigma_0<0$),
is broken by the elastic nonlinearity. Second, the sign of the correction 
is sensitive to whether the kinematic BC at the free surface is important (e.g.~for diffusion of 
extended objects like polymers in a network) or not.

To compare to a real system, we use the following parameter values as suggested by Ref.~\cite{Barbero09}:
$h_{0} = 140 {\rm nm}$ for the thickness of the film and $\bar{E}=5\cdot 10^{5} \mathrm{Pa}$ for the modulus.
For the surface tension we use the value for polystyrene, $\gamma_{PS}\simeq30\cdot10^{-3} {\rm N}{\rm m}^{-1}$. 
Fig.~\ref{ssfig} displays the full growth rates, Eqs.~(\ref{fullss})-(\ref{fullsfkin}) 
in appendix \ref{det}, as a function of reduced wave number $kh_0$.
Note that we renormalized $\bar s(k)=(h_0^3/(M\bar E))s(k)$.
Fig.~\ref{ssfig}a) displays the case of the slip BC at the bottom, 
with (solid curves) and without (dotted curves) 
accounting for the kinematic BC at the free surface.
Finite stresses lead to a Grinfeld instability. Growth rates differ whether compression
($\lambda<0$) or extension ($\lambda>0$) is considered. In case of the fixed BC at the bottom,
see Fig.~\ref{ssfig}b), there exists a threshold stress beyond which the system becomes unstable.
For the chosen surface tension, $\gamma = 0.5 h_{0} \bar{E}$ in reduced units, the system
destabilizes for $\lambda>\lambda_{1,c}\simeq1.387$ and  $\lambda<\lambda_{2,c}\simeq0.426$.
Note that a (symmetric) threshold stress also occurred in the linear model, cf.~Eq.~(\ref{thresh_fixed}).
Fig.~\ref{kmax} displays the dependence of the fastest growing wavenumber on
the prestress $|\sigma_0|/\bar{E}$ as obtained from $\lambda$ by Eq.~(\ref{rel_sig_lambda}).
One clearly sees the asymmetry with respect to compression/stretch for finite stresses.

To summarize, in the last two sections we proposed a general framework that includes the Grinfeld instability
as well as possible buckling. 
The possibility of buckling is due to the coupling 
of surface undulations and the displacement field via a kinemtic BC at the free surface.
One gets corrections to the Grinfeld instability, as contained in the full growth rates given in
appendix \ref{det}.
However, in the small wave-number limit and for thin films, the leading order 
terms are identical with the classical result. For finite stresses the $\pm$-symmetry 
with respect to stresses predicetd by the linear Grinfeld-theory is no longer valid.
In the next section we use the developed framework to study the simultaneous action of 
in-plane residual stress and an electric field, both acting as destabilizing factors
for elastic films.

\section{Addition of external electric field} 
\label{grinE}

Recently the instability of polymeric liquids \cite{SteinerEPL,Barbero09} and elastomers \cite{Sharma08}
in an external electric field acting normal to the film surface has been investigated
experimentally. In Ref.~\cite{Barbero09}, it has been found that the instability is faster
for freshly spin-casted films than for aged films. This suggests that stresses in
the fresh films due to the nonequilibrium production process may be involved
in the destabilization. 
In view of this we generalize the developed approach to the case where 
an external electric field is acting normal to the surface, 
in addition to the stress 
in $x$-direction. 
The electrostatic part will be closely related to previous studies of
elastic instabilities \cite{SharmaPRL,monch2001,Sharmalong,Sharma08} 
due to forces normal to the surface (Van der Waals or electric field).
Related studies have been undertaken in Refs.~\cite{Yang05,Yang06}.
However, there the thin film was regarded as conductive, the external stress
was imposed externally (implying that the base state with applied field
was fixed at $\sigma_{xx}^0=\sigma_0$ rather than $\sigma_{xx}^0=\sigma_0+F$ as 
in our case with $F$ the additional contribution from the electric field, see below)
and the kinematic BC ({\it i.e.} the coupling of film height and displacement field) 
at the free surface was not taken into account.

\begin{figure}[t]
\centering
\vspace{.2cm}
\includegraphics[width=0.48\textwidth]{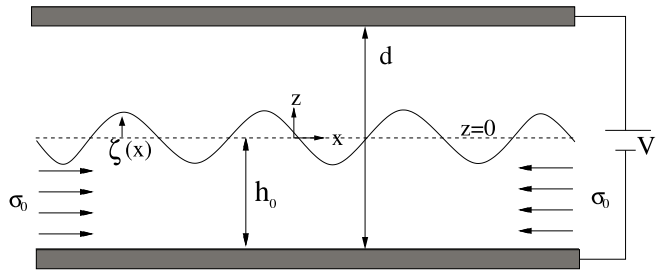}
\caption{\label{sketchE} Sketch of the capacitor geometry. 
There is uniaxial stress in $x$-direction (which could also be tensile). 
An external electric field in $z$-direction (for the unperturbed film)
is imposed by externally applying a voltage $V$ between the electrodes of distance $d$.}
\end{figure}

Let us assume that the polymer film 
is brought into a parallel plate capacitor, see the sketch in Fig.~\ref{sketchE}. 
A voltage difference $V$ is applied over the distance
of the two plates $d$ (the lower plate is at $z=-h_0$, the upper one at $z=d-h_0$).
The gap may be filled with any dielectric.
In view of the experiments in Ref.~\cite{Barbero09}, we take $\epsilon_1\simeq2.5$ 
(polystyrene) as the dielectric constant of the polymer film 
and $\epsilon_2=1$, i.e. the gap is filled with air.

The electric field will introduce a stress at the polymer-air interface.
Let us introduce the Maxwell stress tensor  
\beq
T^{(k)}_{ij}=\epsilon_k \epsilon_0 \left( E^{(k)}_i E^{(k)}_j-\dfrac{1}{2} \left(E^{(k)}\right)^2
\delta_{ij}\right)\,,
\eeq
where the index $k=1$ denotes the polymer film and $k=2$ the gap.
The BC at the free surface, cf. Eq.~(\ref{BCgrin}), now reads
\beq
\label{Max_tensor}
\bn\cdot\left( \sigma^{(1)}- \sigma^{(2)}+T^{(1)} - T^{(2)} \right)\cdot\bn 
=\gamma\p_{xx}h\,, 
\eeq
where we wrote only the linear order expression for the surface tension.
We can put $\sigma^{(2)}=0$ (or to a constant pressure value that is not important),
$\sigma^{(1)}=\sigma$
and define an electrostatic 'pressure' (strictly speaking a normal stress) by 
\beq\label{pEdef}
p_E(h)=\bn\cdot\left(T^{(1)}- T^{(2)}\right)\cdot\bn\,.
\eeq
Note that this electrostatic stress depends on the film thickness, see below.
The BC finally reads
$\bn\cdot \sigma\cdot \bn= \gamma \partial_x^2 h - p_E(h)$.

We now have to evaluate the additional contribution from the electric field.
We can again solve the problem by a perturbative method by writing 
$\mathbf{E}^{(i)}=\mathbf{E}^{(i)}_{0}+\mathbf{E}^{(i)}_{1}$, where 
$\mathbf{E}^{(i)}_{1}$ is the first order correction due to undulations. 
To lowest order, one has to satisfy that the normal dielectric displacement is continuous,
$\epsilon_1 E^{(1)}_{0,z}=E^{(2)}_{0,z}$.
Second, we have $V=h_0 E^{(1)}_{0,z}+(d-h_0)E^{(2)}_{0,z}$, $E^{(i)}_{0,x}=0$ and thus one gets
$E^{(1)}_{0,z}=\epsilon_2 V/(h_0+\left(d-h_0 \right) \epsilon_1)$,
$E^{(2)}_{0,z}=\frac{\epsilon_1}{\epsilon_2} E^{(1)}_{0,z}$.
In the next order, we have to solve Maxwell's equations
\beq\label{Maxwell}
\partial_z  E_{1,x}^{(i)} - \partial_x  E_{1,z}^{(i)}=0\,,\,\,\, 
\partial_z  E_{1,z}^{(i)} + \partial_x  E_{1,x}^{(i)}=0\,,
\eeq
with the BCs 
\beq
E_{1,x}^{(1)}(z=-h_0)=0\,&,&\,\,\,E_{1,x}^{(2)}(z=d-h_0)=0\,,\nonumber\\
\bn\cdot(\epsilon_2 \mathbf{E}^{(2)}-\epsilon_1 \mathbf{E}^{(1)})=0
\,&,&\,\,\,\bt\cdot(\mathbf{E}^{(2)}-\mathbf{E}^{(1)})=0\,.
\eeq
These BC state that the field has to be perpendicular to the conductive electrodes and
that at the film surface one has continuity in the normal displacement and the tangential field.
Assuming $E^{(i)}_{1,z}\propto \cos(k x)$, the system is readily solved yielding
the field components given by Eqs.~(\ref{Ecomp}) in appendix \ref{det}, 
in agreement with Ref.~\cite{OnukiPA1995}.

Evaluating the normal-normal component of the Maxwell stress, 
for the electrostatic pressure as defined in Eq.~(\ref{pEdef}) above
we get to leading order
\beq
-p_E(\zeta)&=& F + Y(k)\,\zeta\,,
\eeq
where
\beq
\label{Fdef}
F&=&-p_E(0)=\frac{1}{2}\frac{\epsilon_0 \epsilon_1 \epsilon_2 (\epsilon_1-\epsilon_2)V^2}
{\left(\epsilon_2 h_0 + \left(d-h_0\right) \epsilon_1 \right)^2}\,,\,\\
\label{Ydef}
Y(k)&=&\frac{-2 k p_E(0)(\epsilon_1-\epsilon_2)}{\left[\epsilon_1 \tanh\left( \left(d-h_0\right) k\right)+\epsilon_2 \tanh\left(h_0 k\right)\right]}\,.
\eeq
Note that both $F$ and $Y$ are strictly positive, $F, Y>0$.
As one has  
$\bn\cdot (T^{(1)}-T^{(2)})\cdot\bt=0$, the tangential BC at the free surface is unchanged by the electric field.
In the base state, the contribution of the electric field will be
an isotropic pressure \cite{SharmaPRL,Sharmalong}, given by $F=-p_E(\zeta=0)$.    
Concerning the displacements relative to the base state, the procedure is completely
analogous to the one in the previous sections.
Only the BC at the free surface, 
and the chemical potential have to be changed
accordingly to include the electric stresses. 
In the chemical potential, Eq.~(\ref{muwithgam}), we have to add the contribution due to 
the electric stress by writing
\beq\label{potchi}
\delta\mu&=&\delta E_{\rm el}-\gamma\kappa+p_E(\zeta)\nonumber\\
&=&\delta E_{\rm el}+\gamma k^2 \zeta -F-Y(k) \zeta\,.
\eeq
As $F$ is a constant, its contribution to surface diffusion vanishes.

The general solution for the displacement field, Eq.~(\ref{gensol_uz}), 
with the BCs at the substrate already imposed, is still valid.
One only has to determine the coefficients 
fulfilling the new BC at the free interface.
These coefficients,
$B_{1,s}^{E}$, $B_{2,s}^{E}$ and $A_{1,f}^{E}$, $B_{1,f}^{E}$ can be obtained
from the respective solutions without field
by the simple substitution 
\beq
\gamma \rightarrow \left(\gamma - \frac{Y(k)}{k^2}\right)\,.
\eeq
This rescaling of $\gamma$ permits 
to obtain also the functions $Z_{s}^{E}(k)$, $Z_{f}^{E}(k)$ 
that determine the stability of the flat base state in the presence
of a field, as well as the growth
rates $s_{s}^{E}$ and $s_{f}^{E}$. 
The obtained expressions are very general. Although unsightly they
contain the physics of buckling, the elasto-electric instability, the 
Grinfeld-instability and surface diffusion in an applied electric field.

\begin{figure}[t]
\centering
\vspace{.2cm}
\includegraphics[width=0.48\textwidth]{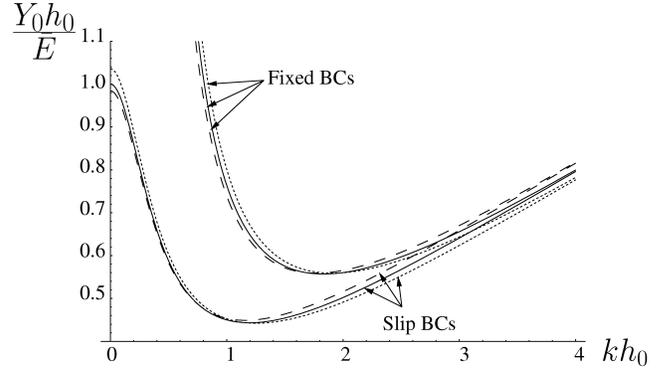}
\caption{\label{Y_instab} Stability diagram for the electric field-induced instability.
The curves display the electric contribution to stress in reduced units, 
$Y_0 h_0/\bar E$, as a function of reduced wave number $h_0 k$
for the slip and the fixed BC at the bottom, as indicated.  
Solid lines correspond to the stress-free case, $\lambda =1.0$.
Dashed lines are for finite stretch, $\lambda =1.1 $, 
and dotted lines for finite compression $\lambda=0.9$.
Parameters: $\gamma = 0.5 h_{0} \bar{E}$.
}
\end{figure}

Let us first discuss the case without surface diffusion.
One expects to get an instability for $Y>Y_c$ as described by Sharma {\it et al.} \cite{SharmaPRL}. 
Note, however, that in case of an applied electric field $Y(k)$ is $k$-dependent, while 
in Ref.~\cite{SharmaPRL} a Van der Waals-interaction with a contactor was studied, 
where $Y$ is a constant.
To compare we write $Y\left(k\right)$ as a function of $Y_{0}$ where 
$Y_0 = \lim\limits_{\substack{k \to 0}} Y\left(k\right)$,
\beq
Y\left(k\right) = \frac{k Y_0 (d \epsilon_1+h_0 (\epsilon_2-\epsilon_1))}
{\epsilon_1 \tanh (k (d-h_0))+\epsilon_2 \tanh (h_0 k)}\,.
\eeq
One gets the following conditions for instabilities
\beq
\label{Y_slip}
Y_0 &=& \dfrac{- Z_s\left(k\right) \left(\epsilon_1 \tanh\left[k \left(d - h_0\right)\right]
+ \epsilon_2 \tanh\left[k h_0\right]\right)}{\lambda k^2 \left(l^2-k^2\right) 
\sinh\left[k h_0\right]\sinh\left[l h_0\right] 
\left(\epsilon_1 d + \left(\epsilon_1 - \epsilon_2 \right)h_0\right)}, \nonumber\\
&&\\
\label{Y_fix}
Y_0&=& \dfrac{2 Z_f\left(k\right) \left(\epsilon_1 \tanh\left[k \left(d - h_0\right)\right] 
+ \epsilon_2 \tanh\left[k h_0\right]\right)}{\lambda k^2 \left(l^2-k^2\right) g^{-}\left(k,l\right)
\left(\epsilon_1 d + \left(\epsilon_1 - \epsilon_2 \right)h_0\right)}
\eeq
for the slip and the fixed BC, respectively.

Fig.~\ref{Y_instab} represents stability diagrams for the elasto-electric instability,
for both BCs as given by Eqs.~(\ref{Y_slip}), (\ref{Y_fix}). For high enough $Y_0$, there exist
solutions with finite $k$. However, as $Y(k)$ depends on $k$ and one has a complicated
dependence on both $\lambda$ and $\gamma$, we could not obtain simple formulas for the threshold.  
For $Y\left(k\right) = {\rm const}$ (as for Van der Waals-interactions) 
we find the same result as given by Ref.~\cite{Pan09} (slip BC) 
and as given in Ref.~\cite{Sharma08,HeEPL} (fixed BC).
We can observe the following general trends due to finite stretches: 
considering the small wave number branch,
to get an undulation with the same small wave number, 
$Y_{0} (\lambda = 0.9) >Y_{0} (\lambda = 1.0) >Y_{0} (\lambda = 1.1) $.
Thus compression acts stabilizing and tension destabilizing on small wave numbers. 
On the other hand, for the large wave number branch one has
$Y_{0} (\lambda = 0.9) <Y_{0} (\lambda = 1.0) <Y_{0} (\lambda = 1.1) $, 
thus tension is stabilizing and compression destabilizing.

Let us now look at the case with surface diffusion.
In the thin film limit $h_0k\ll1$ one gets for the slip BC  
\beq\label{ss_lambda}
s^{E}_{s} (k) &\simeq& s_s(k)+ M k^2 Y_0 \dfrac{\left( 3 \lambda^3 + 1 \right) }
{2 (1+\lambda^3)} \nonumber \\
&&+ M k^4 \dfrac{Y_0h_0\lambda^3/3}{2 \left(\lambda^3 + 1 \right)+(3 \lambda^3 + 1)Y_2},\\
\label{ss_kin_lambda}
s^{E}_{s,kin} (k) &\simeq& M k^2 \dfrac{G}{h_0 \lambda} 
\dfrac{G \left(\lambda^3-1 \right)^2 + h_0 Y_0 \lambda \left( 1 + 3 \lambda^2 \right) }
{G \left(3 + \lambda^3 \right) - h_0 Y_0 \lambda},\qquad   
\eeq
excluding and including the effects of the kinematic BC, respectively.
In the latter case we show only the leading order contribution in $k^2$.
For the fixed BC one gets
\beq
\label{sf_lambda}
s^{E}_{f} (k) &\simeq& s_{f}(k) +M k^2 Y_0 \nonumber\\
&&\hspace{-1cm}+ M k^4 \left(Y_0 \dfrac{h_0^2 \left(\lambda^3-1\right)}{2} +  Y_2 \right), \,\,\,\\  
\label{sf_kin_lambda}
s^{E}_{f,kin} (k) &\simeq& s_{f,kin} (k) + M k^2 Y_0 \nonumber\\
&& \hspace{-1cm}+ M k^4 \left( Y_0 h_0^2 \left(\lambda^3-1\right) + Y_0^2 \dfrac{h_0^3 \lambda}{3 G } + Y_2 \right)\,,
\eeq
where we introduced $Y_2=\frac{1}{2}\frac{d^2}{dk^2}Y(k)_{|k=0}$.

In the limit of small stresses, 
$\lambda=1\pm\delta$ with $\delta=\sigma_0/(3G)\ll 1$,
to lowest order $k^2$ and up to third order in stress one gets 
\beq
\label{ss_sigma}
s^{E}_{s} (k) &\simeq& M k^2 
\bigg[  Y_0  + \sigma_0 \dfrac{Y_0}{\bar{E}}  \nonumber\\
&&\quad\quad\quad +  \sigma_0^2 \dfrac{3 \bar{E}-2\bar{Y_0}}{3 \bar{E}^2 h_0}
- 2  \sigma_0^3 \dfrac{9 \bar{E}+10\bar{Y_0}}{27\bar{E}h_0}  \bigg],\\
\label{ss_kin_sigma}
s^{E}_{s,kin} (k) &\simeq& M k^2 
\bigg[ Y_0 \dfrac{\bar{E}}{\bar{E}-\bar{Y_0}} + 
\sigma_0 Y_0 \dfrac{6\bar{E}-5\bar{Y_0}}{3 \left(\bar{E}-\bar{Y_0}\right)^2}\nonumber\\
&&\hspace{-1cm} + \sigma_0^2 \frac{9\bar{E}^3-12 \bar{E}^2 \bar{Y_0}-12 \bar{E} \bar{Y_0}^2 +16 \bar{Y_0}^3 }
{9 \bar{E} h_0 \left( \bar{E}-\bar{Y_0}\right)^3}\nonumber\\
&&\hspace{-1cm} + \sigma_0^3 \frac{9\bar{E}^4-76 \bar{E}^3 \bar{Y_0}+116 \bar{E}^2 \bar{Y_0}^2-64 
\bar{E} \bar{Y_0}^3 +16 \bar{Y_0}^4 }
{27 \bar{E}^2 h_0 \left( \bar{E}-\bar{Y_0}\right)^4}\bigg],\nonumber\\
\eeq
where we introduced $\bar{Y_0}=h_0Y_0$, and
\beq
\label{sf_kin_sigma}
s^{E}_{f} (k) = s^{E}_{f,kin} (k)&\simeq& M k^2 Y_0\,. 
\eeq

\begin{figure}[t]
\centering
\vspace{.2cm}
\includegraphics[width=0.45\textwidth]{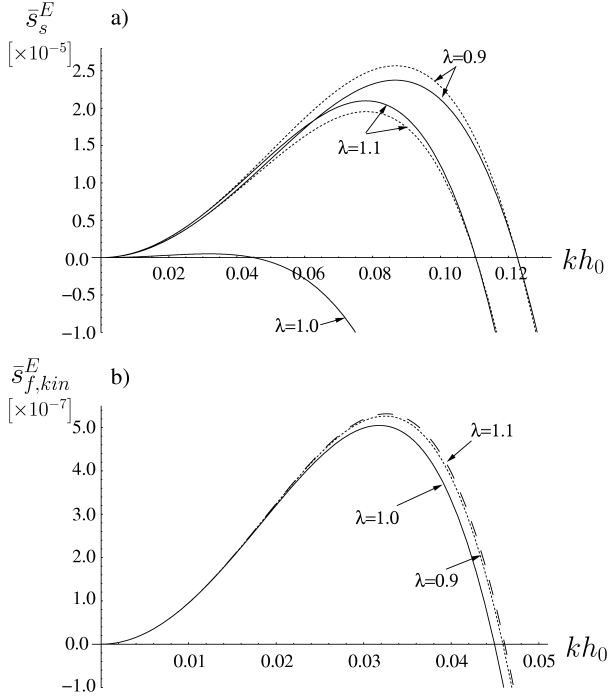}   
\caption{\label{Growth_rate} 
Growth rates as a function of reduced wave number 
for different stretch factors $\lambda$, corresponding to different prestress, and with finite voltage applied
normal to the film. 
Panel a) displays the case of the slip BC at the bottom. Without stress, $\lambda=1$, 
the system is slightly unstable due to field-induced diffusion.
Compression ($\lambda<1$) or stretch ($\lambda>1$) lead to a more pronounced instability, 
but with differing rates.
Solid lines have been obtained with the kinematic BC at the free surface, 
pointed only with surface diffusion.
Panel b) shows the case of the fixed BC at the bottom, where finite 
stresses only lead to small corrections.
Parameters: $\gamma = 0.5 h_{0} \bar{E}$, $\epsilon_1=2.5$, $\epsilon_2=1$, $d/h_0=5$, $h_0Y_0=0.001\bar E$.
}
\end{figure}

For the slip BC, Eqs.~(\ref{ss_lambda}, \ref{ss_kin_lambda}) 
display a coupling between prestress and the applied electric stress at the order $\mathcal{O}\left(k^2\right)$.
Hence the application of the field breaks the $\pm\sigma_0$-symmetry
already in lowest order in stress.
In the small stress limit 
this coupling is linear like $Y_0\sigma_0$.
As one usually has $h_0Y_0\ll\bar{E}$ (otherwise the elasto-electric instability takes over), 
for small wavelengths one might be driven to the conclusion
that the coupling between the electric field and the stress like $Y_{0} \sigma_0$
implies that compression acts stabilizing while stretch acts destabilizing.
However, the destabilizing contribution from the electric field, $M k^2 Y_0$, 
is usually dominating  and thus $\bar{s}^{E}>\bar{s}$, 
compare e.g. Figs.~\ref{ssfig}a) and \ref{Growth_rate}a).
Thus the influence of the coupling is observed rather 
beyond the maximum of the growth rate, cf. 
Fig.~\ref{Growth_rate}a),
as a gap between the curves with compression and stretch.
However, in case of $\sigma_0\ll h_0Y\ll\bar E$, the two destabilizing forces
do not add and the growth rate under compression is indead slightly smaller
than the growth rate of an unstressed film. 
As in the case without field one observes finite stress effects yielding 
positive contributions like $+\sigma_0^3$ with the kinetic BC and negative ones 
like $-\sigma_0^3$ with the non-kinetic version. Positive and negative prestress have thus
opposite effects and these effects depend on the kinematic BC.

Fig.~\ref{Growth_rate}a) displays the general growth rates for the slip BC.
For $\lambda<1$ ($\sigma_0<0$), 
the kinematic version yields a smaller growth rate than the non-kinematic version. 
However, for $\lambda>1$ ($\sigma_0 > 0$) the opposite is true.
Fig.~\ref{Growth_rate}b) shows that for the fixed BC finite stress
only leads to higher order corrections, since the leading order destabilization is 
$+k^2 Y_0$ while the stress contributions are proportional to $k^4$. 
Visible differences between $\bar{s}_{f,kin}\left(k\right)$
and $\bar{s}_{f}\left(k\right)$ appear only for rather high stretch factors, namely 
$\lambda\gtrsim1.5$ or $\lambda \lesssim 0.7$ for the chosen surface tension.
For Fig.~\ref{Growth_rate} we used again parameters as suggested by Ref.~\cite{Barbero09}, 
namely a electrode distance of
$d = 5 h_{0}$, dielectric constants $\epsilon_2=2.5$ and $\epsilon_{1}=1$ for the PS film and the air gap, respectively, and
a voltage of $V=16\mathrm{V}$.
In reduced units this leads to $\gamma \simeq 0.5 h_{0} \bar{E}$, $Y_{0} h_{0} \simeq 0.0013 {\bar{E}}$.

Let us briefly discuss the relation of this work to the
experiments of Ref.~\cite{Barbero09}. There it has been found that
freshly produced films, that are supposedly stressed due to the nonequilibrium 
preparation process of spin-coating, have faster growth rates
than aged films -- which had time to relax residual stresses.
This is in accordance with our findings that the two destabilization
mechanisms, the Grinfeld mechanism and the electric force acting on the free surface
of the film, in general join forces. However, in Ref.~\cite{Barbero09} it has 
been found that the wave number of the instability 
is smaller for fresh films than for aged films.
This is in contrast to our calculations, as in the general case 
the unstable wave numbers increase
with stress, see also Fig.~\ref{Growth_rate}.
There are several possible reasons for this discrepancy,
like viscoelastic effects in the film, inhomogeneities, crust formation
due to spin-coating \cite{deGennes02}, etc.
As a next step we plan to generalize the approach proposed here to the 
viscoelastic case, to come closer to these experiments. 

Another interesting point is that the compressive-tensile symmetry 
holding for the stress in case of the Grinfeld instability is broken
in several ways: 
i) by finite stresses, ii) due to the kinematic BC, 
i.e.~the coupling of film height and displacement field, and 
iii) due to the presence of the external electric field.
While the effect is of order $\sigma_0^3$ in the absence of an electric field, 
it is of order $Y_0\sigma_0$ in the presence of field. 
Thus especially in an external electric field, surface undulations 
have noticeably different growth rates and 
this may be used experimentally to determine 
whether stresses in thin films are compressive or tensile. 

\begin{table}
{
\renewcommand{\arraystretch}{1.7}
\begin{tabular}{| l | c | c | c |c|}
\hline
system         &  BC   &           & destabilization $s\simeq$                  
& \hspace{1mm} Eq. \hspace{1mm} \\ \hline\hline
semi-$\infty$ \hspace{1mm} &       &           & $M k^2\frac{\sigma_0^2}{\bar E}k$  
& \hspace{1mm} (\ref{DEhinf}) \hspace{1mm}  \\ \hline
thin film  \hspace{1mm}    & \hspace{1mm}slip \hspace{1mm} &           & $M k^2\frac{\sigma_0^2}{\bar E h_0}$ 
& \hspace{1mm} (\ref{DEhthin}) \hspace{1mm} \\ \hline
thin film    & fixed &           & $M k^2 \frac{4 \sigma_0^2}{\bar E h_0}(h_0k)^2$ 
& \hspace{.2mm} (\ref{DEhthinfix})\hspace{1mm} \\ \hline
thin film     & slip  & \hspace{1mm} el.field \hspace{1mm} & $M k^2\left(Y+\frac{h_0Y\sigma_0+\sigma_0^2}{\bar E h_0}\right)$ 
& \hspace{1mm} (\ref{ss_sigma})     \hspace{1mm}  \\ \hline
thin film     & kin,slip  & \hspace{1mm} el.field \hspace{1mm} & $M k^2\left(Y+\frac{2h_0Y\sigma_0+\sigma_0^2}{\bar E h_0}\right)$ 
& \hspace{1mm} (\ref{ss_kin_sigma})     \hspace{1mm}  \\ \hline
thin film     & fixed & el.field  & $M k^2Y$ 
& \hspace{1mm} (\ref{sf_kin_sigma})  \hspace{1mm} \\ \hline
\end{tabular}
}
\caption{\label{T2} Summary of the leading order destabilization terms 
in the growth rate of surface undulations, $s(k)$, for
different BCs at the bottom and with or without electric field.
Both stress and electric field are assumed small, $\sigma_0,h_0 Y\ll\bar E$.}
\end{table}

\section{Conclusions and perspective}
\label{Concl}

We have studied the instability of a polymer film
under the simultaneous action of internal stress and an
externally applied electric field.
For this purpose we formulated a general framework that
has a very rich phenomenology:
in absence of surface diffusion the system is stable against buckling
but displays an electrically induced instability towards periodic undulations. 
In case that the polymer chains are able to diffuse close to the surface
due to gradients in the chemical potential,
the Grinfeld mechanism becomes active, as well as a destabilizing contribution
induced by the external electric field. The growth rates of surface undulations
are sensitive to the boundary conditions at the bottom
and have a rich phenomenology, see Table \ref{T2}. 

Our approach also highlights the importance of the coupling between the height
of the film's surface and the displacement field inside the film, 
which naturally arises from the kinematic boundary condition at the film surface.
This coupling has been neglected in previous studies.
Its consequences can be seen as finite stress 
corrections to the Grinfeld instability, and analogously for the 
electric instability. Moreover, this coupling establishes the connection between
the above mentioned elasto-electric instabilities and the Grinfeld-like diffusive instabilities,
as becomes apparent from the general growth rates of height fluctuations. 
These growth rates 
have been calculated as a function of internal stress, electric field, mobility of the
chains and surface tension. It is shown that both destabilizing factors,
internal stress and electric field, generally add. 

The relevance for recent experiments on spin-cast
thin polymer films has been only briefly discussed.
A generalization to the viscoelastic case, and possibly also including more structural details 
of spin-cast film, is needed to account for these experiments.
In turn, as the experiments can measure separately
the most unstable wavelength and the growth rate, 
they could give direct access to the internal stress and to 
the mobility of polymers in thin films,
which both are of technological importance.
In particular, as the electric field makes the breakage of the 
compressive-tensile symmetry of the Grinfeld instability induced by the
coupling to the displacement field noticeable,
careful measurements of the growth rates could be used to determine
the nature of the stresses, i.e. whether they are compressive or tensile.

The authors would like to thank Ken Sekimoto for stimulating discussions.

\appendix

\section{Results for stress functions, displacement coefficients, electric field}
\label{det}

For the Airy stress functions in the Grinfeld calculation (cf. section \ref{Grin_elastic}) one gets
\beq\label{chisol}
\chi_s(x,z) \hspace{-1mm}&=&\hspace{-1mm}\eps\cos(kx)
\big[\,a_s\cosh(k(z+h_{0}))\nonumber\\
&&\hspace{1.3cm}+b_s(z+h_{0})\,\sinh(k(z+h_{0}))\big]\quad\,\,
\eeq 
in case of the slip BC,
with
\beq
a_s&=&2A\frac{h_0 k \gamma \cosh\left(h_0 k  \right)+\left( \gamma + \sigma_0 h_0\right)\sinh \left( h_0 k\right) }
{2h_0k+\sinh\left(2 h_0 k\right) }\,,\quad\nonumber\\
\label{asbs}
b_s&=&-2A\frac{k \gamma \sinh\left(h_0 k  \right)+\sigma_0 \cosh \left( h_0 k\right)}
{2h_0k+\sinh\left(2 h_0 k\right) }\,.
\eeq
In case of the fixed BC, one calculates
\beq\label{chisolf}
\chi_f(x,z) \hspace{-1mm}&=&\hspace{-1mm}\eps\cos(kx)
\big[\,a_f\cosh(k(z+h_{0}))\nonumber\\
&&\hspace{1.3cm}+b_f(z+h_{0})\,\cosh(k(z+h_{0})) \nonumber\\
&&\hspace{1.3cm}+a_f k (z+h_0) e^{-k(h_0+z)}\big]\,\,\,\,
\eeq
with
\beq
a_f\hspace{-1mm}&=&\hspace{-1mm}2A \frac{  (h_0 \sigma_0+\gamma )\cosh (h_0 k)+h_0 k \gamma  \sinh (h_0 k)}{2
   h_0^2 k^2+\cosh (2 h_0 k)+1}\,,\quad\nonumber\\
\label{afbf}
b_f\hspace{-1mm}&=&\hspace{-1mm}-2A\frac{(\sigma_0+k\gamma)\cosh(h_0 k)+h_0 k (\sigma_0 - k \gamma )e^{-h_0 k} }
{2 h_0^2 k^2+\cosh (2 h_0 k)+1}\,.\quad\,\,\,\,\,\,\,
\eeq

The full growth rates calculated in section \ref{grindiffuse} read
\begin{widetext}
\beq
\label{fullss}
s_s(k)&=&M \dfrac{2 G \left( l^2-k^2 \right)^3 \cosh\left[h_0 k  \right]\cosh\left[h_0 l  \right]-k^3 \gamma \lambda \left[b\left(k,l\right)+b\left(k,-l\right) \right] }
{2 \lambda \left(k^2 + l^2 \right) g^+\left(k,l\right) }, \\
\label{fullsskin}
s_{s,kin} (k)&=&M l G \dfrac{2 G \left( l^2-k^2 \right)^3 \cosh\left[h_0 k  \right]\cosh\left[h_0 l  \right]-\gamma k^3 \lambda \left[b\left(k,l\right)+b\left(k,-l\right) \right] }
{\lambda \left[2 \gamma k^3 \lambda \left(l^2 - k^2 \right)\sinh\left[h_0 k  \right]\sinh\left[h_0 l  \right] + l G  \left(3 k^2 + l^2 \right) g_{1}\left(k,l\right)
 + k G  \left( k^2 + l^2 \right) g^-\left(k,l\right) \right]}, \\
\label{fullsf}
s_f(k)&=& -M \dfrac{\left( 8 k^4 l^2 \gamma \lambda \left(k^2 + l^2 \right) - k^3 \gamma \lambda \left[r\left(k,l \right)+r\left(k,-l \right) \right] 
+ G \left(l-k\right)^3 \left(l+k\right)^3 g\left(k,l \right)\right)}{2 \lambda \left(k^4 + 6 k^2 l^2 + l^4 - \left(k^2+l^2\right) f\left(k,l\right)\right)}, \\
\label{fullsfkin}
s_{f,kin}(k) &=& -M l G \dfrac{\left( 8 k^4 l^2 \gamma \lambda \left(k^2 + l^2 \right) - k^3 \gamma \lambda \left[r\left(k,l \right)+r\left(k,-l \right) \right] 
+ G \left(l-k\right)^3 \left(l+k\right)^3 g_{1}\left(k,l \right)\right)}{\lambda \left[ \left(l^2-k^2 \right) \lambda \gamma k^3 g^-\left(k,l\right) + G \left(r\left(-l,k\right)-r\left(l,k\right) \right)
+8 G l k^2 \left(l^2+k^2\right) \right]}, 
\eeq
\end{widetext}
where we have introduced
\beq
b(k,l)=(k^4+2k^2 l^2 + 4 k l^3 + l^4 ) \sinh \left[\left(l-k \right) h_0 \right].\quad
\eeq

In section \ref{grinE} the electric field has to be calculated to linear order. From the
Maxwell equations with suitable BC, as given in the main text, one gets
\beq
E_{1,z}^{(1)}&=&\epsilon h\cos(k x) \epsilon_2 \tilde E\, \frac{\cosh\left(k\left(z+h_0\right)\right)}{\cosh\left(h_0 k\right)},\nonumber\\
E_{1,x}^{(1)}&=&-\epsilon h\sin(k x) \epsilon_2 \tilde E\, \frac{\sinh\left(k\left(z+h_0\right)\right)}{\cosh\left(h_0 k\right)},\nonumber\\
E_{1,z}^{(2)}&=&\epsilon h\cos(k x) \epsilon_1 \tilde E\, \frac{\cosh\left(k\left(z+h_0-d\right)\right)}{\cosh\left(k (d-h_0)\right)},\nonumber\\
\label{Ecomp}
E_{1,x}^{(2)}&=&-\epsilon h\sin(k x) \epsilon_1 \tilde E\, \frac{\sinh\left(k\left(z+h_0-d\right)\right)}{\cosh\left(k (d-h_0)\right)},\,\,\,\,
\eeq
with the abbreviation
\beq
\tilde E&=&\frac{V}{(\epsilon_2 h_0+\left(d-h_0\right)\epsilon_1)} \nonumber\\
&& \cdot\,\frac{ k (\epsilon_1-\epsilon_2)}{\epsilon_1 \tanh (k (d-h_0))+\epsilon_2 \tanh (h_0 k)}.
\eeq

\section{Alternative motivation for the kinematic BC}
\label{motkinBC}

\begin{figure}[t!]
\centering
\vspace{.2cm}
\includegraphics[width=0.48\textwidth]{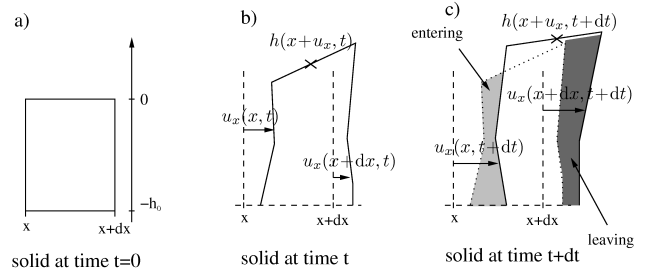}
\caption{\label{figmot} Sketch for motivation for kinematic BC.}
\end{figure}

Here we want to give a more explicit motivation for the kinematic BC with surface diffusion.
Let us consider a small part of an elastic material bounded by $x$ and $x+\mathrm{d}x$ on the $x$-axis,
$h\left(x,0\right)=0$ and $-h_{0}$ on the $z$-axis, see Fig.~\ref{figmot}a).
The extension in $y$-direction, $L_y$, is assumed not to change in time, to stay within 
the plane strain situation.  
At time $t$, the respective bounds are between $x + u_x\left(x,z,t\right)$ and 
$x+\mathrm{d}x+ u_x\left(x+\mathrm{d}x,z,t\right)$, as well as $h\left(x+u_x\left(x,0,t\right),0\right)$ 
and $-h_{0}$, see Fig.~\ref{figmot}b).
Let us use this state, excluding the upper free surface, as a control state and calculate its evolution in time.

The volume inside the considered piece of material is 
\beq
V(t)&=& h\left(x+u_x\left(x,0,t\right),t\right) L_{y}\nonumber \\ 
& & \left(x+\mathrm{dx}+u_x\left(x+\mathrm{dx},z,t\right) -x-u_x\left(x,z,t\right)\right),\nonumber\\
V(t+\mathrm{d}t)&=& h\left(x+u_x\left(x,0,t+\mathrm{dt}\right),t+\mathrm{dt}\right) L_{y} \nonumber \\
& & \left(x+\mathrm{dx}+u_x\left(x+\mathrm{dx},z,t\right) -x-u_x\left(x,z,t\right)\right)\nonumber
\eeq
at times $t$ and $t+dt$, respectively.
The volume change $\delta V = V(t+\mathrm{d}t) - V(t)$ thus reads
\beq\label{deltaV1}
\delta V &=& \partial_t \left( h + u_x\left(x,0,t\right)\partial_x h \right)\mathrm{dt} L_{y} \mathrm{dx} \left( 1 + \partial_x u_x \right) \nonumber \\
&\approx& \partial_t h \, \mathrm{dt} \, L_{y} \, \mathrm{dx}\,,
\eeq
at first order $\mathcal{O}(\epsilon)$. As $h(x,t)$ can be any function, this is true also if there
are additional surface processes.

On the other hand, we can express the change in volume by 
changes due to displacements and surface diffusion on
both boundaries.
During an infinitesimal time $\mathrm{d}t$, the volume change due displacements on the left boundary  
$L(z,t)=x+u_x\left(x,z,t\right) $ is 
\beq
\simeq \int_{-h_0}^{h\left(x,t\right)}\hspace{-6mm} \left[ u_x\left(x,z,t+\mathrm{dt}\right) 
- u_x\left(x,z,t\right) \right] \mathrm{d}z L_y 
\eeq
at first order $\mathcal{O}(\epsilon)$ (which allowed us to replace the upper integral bound 
$h\left(L(z,t),t\right)$ by $h\left(x,t\right)$).
The volume entering by diffusion along the free surface on $L(z=0,t)$ is
\beq
J_{{\rm surf},x}\left(x+u_x\left(x,t\right),t\right) \mathrm{dt} L_{y} \simeq J_{{\rm surf},x}\left(x,t\right) \mathrm{dt} L_{y}.\quad
\eeq
During the same time $\mathrm{d}t$, at the 
right boundary $R(z,t)=x+\mathrm{dx}+u_x\left(x+\mathrm{dx},z,t\right)$ we get
\beq
\simeq 
\int_{-h_0}^{h\left(x+\mathrm{dx},t\right)}\hspace{-12mm}
\left[ u_x\left(x+\mathrm{dx},z,t+\mathrm{dt}\right) - u_x\left(x+\mathrm{dx},z,t\right) \right] \mathrm{d}z L_{y},\qquad\,\,\,
\eeq
where again we simplified the integral bound, $h\left(R(z,t),t\right)$
to $h(x+\mathrm{dx},t)$.
Surface diffusion at the right boundary contributes 
\beq
J_{{\rm surf},x}\left(x+\mathrm{dx}+u_x\left(x+\mathrm{dx},t\right),t\right) \mathrm{dt} L_{y} \nonumber \\ 
\simeq J_{{\rm surf},x}\left(x+\mathrm{dx},t\right) \mathrm{dt} L_{y}\,.
\eeq
From $\delta V=$(change at $R$) $-$ (change at $L$) we obtain
\beq\label{deltaV2}
\delta V &\approx& -\partial_x J_{{\rm surf},x}\,\mathrm{dt}\,\mathrm{dx}\,L_{y}\nonumber \\
& & - \mathrm{dx}\,L_y\,\mathrm{dt}\,\partial_t \left(\displaystyle \int_{-h_0}^{h\left(x,t\right)} 
\partial_x u_x\left(x,z,t\right)\mathrm{dz}\right).\,\,\,
\eeq
Using the incompresibility condition, integrated over the film thickness, one gets
\beq
& &\int_{-h_0}^{h\left(x,t\right)} \partial_x u_x\left(x,z,t\right) \mathrm{d}z 
\approx - u_z\left(x,0,t\right),\nonumber
\eeq
where we used that $u_z=0$ holds at the bottom.
With the surface flux given by
$J_{{\rm surf},x}\left(x,t\right) = - M \partial_x \left(\delta \mu \right)$,
equality of Eq.(\ref{deltaV1}) and Eq.(\ref{deltaV2}) implies
\beq
\partial_t h&=&\partial_t u_{z|z=0}+M\partial_x^2\left(\delta \mu_{|z=0}\right)\,.
\eeq
which is Eq.~(\ref{surdiffus}) at first order  $\mathcal{O}(\epsilon)$.


\begin{thebibliography}{38}
\expandafter\ifx\csname natexlab\endcsname\relax\def\natexlab#1{#1}\fi
\expandafter\ifx\csname bibnamefont\endcsname\relax
  \def\bibnamefont#1{#1}\fi
\expandafter\ifx\csname bibfnamefont\endcsname\relax
  \def\bibfnamefont#1{#1}\fi
\expandafter\ifx\csname citenamefont\endcsname\relax
  \def\citenamefont#1{#1}\fi
\expandafter\ifx\csname url\endcsname\relax
  \def\url#1{\texttt{#1}}\fi
\expandafter\ifx\csname urlprefix\endcsname\relax\def\urlprefix{URL }\fi
\providecommand{\bibinfo}[2]{#2}
\providecommand{\eprint}[2][]{\url{#2}}

\bibitem[{\citenamefont{Croll}(1979)}]{Croll1979}
\bibinfo{author}{\bibfnamefont{S.~G.} \bibnamefont{Croll}},
  \bibinfo{journal}{J. Appl. Polym. Sci.} \textbf{\bibinfo{volume}{23}},
  \bibinfo{pages}{847} (\bibinfo{year}{1979}).

\bibitem[{\citenamefont{Reiter et~al.}(2005)\citenamefont{Reiter, Hamieh,
  Damman, Sclavons, Gabriele, Vilmin, and Rapha\"el}}]{Reiter:2005}
\bibinfo{author}{\bibfnamefont{G.}~\bibnamefont{Reiter}},
  \bibinfo{author}{\bibfnamefont{M.}~\bibnamefont{Hamieh}},
  \bibinfo{author}{\bibfnamefont{P.}~\bibnamefont{Damman}},
  \bibinfo{author}{\bibfnamefont{S.}~\bibnamefont{Sclavons}},
  \bibinfo{author}{\bibfnamefont{S.}~\bibnamefont{Gabriele}},
  \bibinfo{author}{\bibfnamefont{T.}~\bibnamefont{Vilmin}}, \bibnamefont{and}
  \bibinfo{author}{\bibfnamefont{E.}~\bibnamefont{Rapha\"el}},
  \bibinfo{journal}{Nature Mat.} \textbf{\bibinfo{volume}{4}},
  \bibinfo{pages}{754} (\bibinfo{year}{2005}).

\bibitem[{\citenamefont{Bodiguel and Fretigny}(2006)}]{Fretigny}
\bibinfo{author}{\bibfnamefont{H.}~\bibnamefont{Bodiguel}} \bibnamefont{and}
  \bibinfo{author}{\bibfnamefont{C.}~\bibnamefont{Fretigny}},
  \bibinfo{journal}{Eur. Phys. J. E} \textbf{\bibinfo{volume}{19}},
  \bibinfo{pages}{185} (\bibinfo{year}{2006}).

\bibitem[{\citenamefont{Vilmin and
  Rapha\"el}(2006{\natexlab{a}})}]{Raphael:2006.1}
\bibinfo{author}{\bibfnamefont{T.}~\bibnamefont{Vilmin}} \bibnamefont{and}
  \bibinfo{author}{\bibfnamefont{E.}~\bibnamefont{Rapha\"el}},
  \bibinfo{journal}{Eur. Phys. J. E} \textbf{\bibinfo{volume}{21}},
  \bibinfo{pages}{161} (\bibinfo{year}{2006}{\natexlab{a}}).

\bibitem[{\citenamefont{Damman et~al.}(2007)\citenamefont{Damman, Gabriele,
  Coppee, Desprez, Villers, Vilmin, Rapha\"el, Hamieh, {Al Akhrass}, and
  Reiter}}]{Reiter:2007.2}
\bibinfo{author}{\bibfnamefont{P.}~\bibnamefont{Damman}},
  \bibinfo{author}{\bibfnamefont{S.}~\bibnamefont{Gabriele}},
  \bibinfo{author}{\bibfnamefont{S.}~\bibnamefont{Coppee}},
  \bibinfo{author}{\bibfnamefont{S.}~\bibnamefont{Desprez}},
  \bibinfo{author}{\bibfnamefont{D.}~\bibnamefont{Villers}},
  \bibinfo{author}{\bibfnamefont{T.}~\bibnamefont{Vilmin}},
  \bibinfo{author}{\bibfnamefont{E.}~\bibnamefont{Rapha\"el}},
  \bibinfo{author}{\bibfnamefont{M.}~\bibnamefont{Hamieh}},
  \bibinfo{author}{\bibfnamefont{S.}~\bibnamefont{{Al Akhrass}}},
  \bibnamefont{and} \bibinfo{author}{\bibfnamefont{G.}~\bibnamefont{Reiter}},
  \bibinfo{journal}{Phys. Rev. Lett.} \textbf{\bibinfo{volume}{99}},
  \bibinfo{pages}{036101} (\bibinfo{year}{2007}).

\bibitem[{\citenamefont{Ziebert and Rapha\"el}(2009)}]{FZER1}
\bibinfo{author}{\bibfnamefont{F.}~\bibnamefont{Ziebert}} \bibnamefont{and}
  \bibinfo{author}{\bibfnamefont{E.}~\bibnamefont{Rapha\"el}},
  \bibinfo{journal}{Phys. Rev. E} \textbf{\bibinfo{volume}{79}},
  \bibinfo{pages}{031605} (\bibinfo{year}{2009}).

\bibitem[{\citenamefont{Vilmin and
  Rapha\"el}(2006{\natexlab{b}})}]{Raphael:2006.3}
\bibinfo{author}{\bibfnamefont{T.}~\bibnamefont{Vilmin}} \bibnamefont{and}
  \bibinfo{author}{\bibfnamefont{E.}~\bibnamefont{Rapha\"el}},
  \bibinfo{journal}{Phys. Rev. Lett.} \textbf{\bibinfo{volume}{97}},
  \bibinfo{pages}{036105} (\bibinfo{year}{2006}{\natexlab{b}}).

\bibitem[{\citenamefont{Asaro and Tiller}(1972)}]{asaro:72}
\bibinfo{author}{\bibfnamefont{R.}~\bibnamefont{Asaro}} \bibnamefont{and}
  \bibinfo{author}{\bibfnamefont{W.}~\bibnamefont{Tiller}},
  \bibinfo{journal}{Metall. Trans. A} \textbf{\bibinfo{volume}{3}},
  \bibinfo{pages}{1789} (\bibinfo{year}{1972}).

\bibitem[{\citenamefont{Grinfeld}(1986)}]{grinfeld:86}
\bibinfo{author}{\bibfnamefont{M.}~\bibnamefont{Grinfeld}},
  \bibinfo{journal}{Sov. Phys. Dokl.} \textbf{\bibinfo{volume}{31}},
  \bibinfo{pages}{831} (\bibinfo{year}{1986}).

\bibitem[{\citenamefont{Grinfeld}(1993)}]{grinfeld:93}
\bibinfo{author}{\bibfnamefont{M.}~\bibnamefont{Grinfeld}},
  \bibinfo{journal}{J. Nonlinear Sci.} \textbf{\bibinfo{volume}{3}},
  \bibinfo{pages}{35} (\bibinfo{year}{1993}).

\bibitem[{\citenamefont{Shenoy and Sharma}(2001)}]{SharmaPRL}
\bibinfo{author}{\bibfnamefont{V.}~\bibnamefont{Shenoy}} \bibnamefont{and}
  \bibinfo{author}{\bibfnamefont{A.}~\bibnamefont{Sharma}},
  \bibinfo{journal}{Phys. Rev. Lett.} \textbf{\bibinfo{volume}{86}},
  \bibinfo{pages}{119} (\bibinfo{year}{2001}).

\bibitem[{\citenamefont{M\"onch and Herminghaus}(2001)}]{monch2001}
\bibinfo{author}{\bibfnamefont{W.}~\bibnamefont{M\"onch}} \bibnamefont{and}
  \bibinfo{author}{\bibfnamefont{S.}~\bibnamefont{Herminghaus}},
  \bibinfo{journal}{Europhys. Lett.} \textbf{\bibinfo{volume}{53}},
  \bibinfo{pages}{525} (\bibinfo{year}{2001}).

\bibitem[{\citenamefont{Shenoy and Sharma}(2002)}]{Sharmalong}
\bibinfo{author}{\bibfnamefont{V.}~\bibnamefont{Shenoy}} \bibnamefont{and}
  \bibinfo{author}{\bibfnamefont{A.}~\bibnamefont{Sharma}},
  \bibinfo{journal}{J. Mech. Phys. Solids} \textbf{\bibinfo{volume}{50}},
  \bibinfo{pages}{1155} (\bibinfo{year}{2002}).

\bibitem[{\citenamefont{Sarkar et~al.}(2008)\citenamefont{Sarkar, Sharma, and
  Shenoy}}]{Sharma08}
\bibinfo{author}{\bibfnamefont{J.}~\bibnamefont{Sarkar}},
  \bibinfo{author}{\bibfnamefont{A.}~\bibnamefont{Sharma}}, \bibnamefont{and}
  \bibinfo{author}{\bibfnamefont{V.}~\bibnamefont{Shenoy}},
  \bibinfo{journal}{Phys. Rev. E} \textbf{\bibinfo{volume}{77}},
  \bibinfo{pages}{031604} (\bibinfo{year}{2008}).

\bibitem[{\citenamefont{Barbero and Steiner}(2009)}]{Barbero09}
\bibinfo{author}{\bibfnamefont{D.~R.} \bibnamefont{Barbero}} \bibnamefont{and}
  \bibinfo{author}{\bibfnamefont{U.}~\bibnamefont{Steiner}},
  \bibinfo{journal}{Phys. Rev. Lett.} \textbf{\bibinfo{volume}{102}},
  \bibinfo{pages}{248303} (\bibinfo{year}{2009}).

\bibitem[{\citenamefont{Sch\"affer et~al.}(2001)\citenamefont{Sch\"affer,
  Thurn-Albrecht, Russell, and Steiner}}]{SteinerEPL}
\bibinfo{author}{\bibfnamefont{E.}~\bibnamefont{Sch\"affer}},
  \bibinfo{author}{\bibfnamefont{T.}~\bibnamefont{Thurn-Albrecht}},
  \bibinfo{author}{\bibfnamefont{T.~P.} \bibnamefont{Russell}},
  \bibnamefont{and} \bibinfo{author}{\bibfnamefont{U.}~\bibnamefont{Steiner}},
  \bibinfo{journal}{Europhys. Lett.} \textbf{\bibinfo{volume}{53}},
  \bibinfo{pages}{518} (\bibinfo{year}{2001}).

\bibitem[{Noz()}]{Nozbook}
\bibinfo{howpublished}{P.~Nozi{\`e}res, in {\it Solids far from Equilibrium},
  edited by C. Godr\`eche (Cambridge University Press, Cambridge, 1992)}.

\bibitem[{\citenamefont{Cantat et~al.}(1998)\citenamefont{Cantat, Kassner,
  Misbah, and M{\"u}ller-Krumbhaar}}]{cantat:98}
\bibinfo{author}{\bibfnamefont{I.}~\bibnamefont{Cantat}},
  \bibinfo{author}{\bibfnamefont{K.}~\bibnamefont{Kassner}},
  \bibinfo{author}{\bibfnamefont{C.}~\bibnamefont{Misbah}}, \bibnamefont{and}
  \bibinfo{author}{\bibfnamefont{H.}~\bibnamefont{M{\"u}ller-Krumbhaar}},
  \bibinfo{journal}{Phys. Rev. E} \textbf{\bibinfo{volume}{58}},
  \bibinfo{pages}{6027} (\bibinfo{year}{1998}).

\bibitem[{\citenamefont{Maugis}(2000)}]{Maugis}
\bibinfo{author}{\bibfnamefont{D.}~\bibnamefont{Maugis}},
  \emph{\bibinfo{title}{Contact, Adhesion and Rupture of Elastic Solids}}
  (\bibinfo{publisher}{Springer}, \bibinfo{address}{Berlin Heidelberg},
  \bibinfo{year}{2000}).

\bibitem[{\citenamefont{Landau and Lifshitz}(1986)}]{landau_el}
\bibinfo{author}{\bibfnamefont{L.~D.} \bibnamefont{Landau}} \bibnamefont{and}
  \bibinfo{author}{\bibfnamefont{E.~M.} \bibnamefont{Lifshitz}},
  \emph{\bibinfo{title}{Theory of Elasticity}} (\bibinfo{publisher}{Pergamon
  Press}, \bibinfo{address}{New York}, \bibinfo{year}{1986}).

\bibitem[{\citenamefont{Nozi{\`e}res}(1993)}]{nozieres:93}
\bibinfo{author}{\bibfnamefont{P.}~\bibnamefont{Nozi{\`e}res}},
  \bibinfo{journal}{J. Phys. I France} \textbf{\bibinfo{volume}{3}},
  \bibinfo{pages}{681} (\bibinfo{year}{1993}).

\bibitem[{\citenamefont{Kohlert et~al.}(2003)\citenamefont{Kohlert, Kassner,
  and Misbah}}]{Kohlert:03}
\bibinfo{author}{\bibfnamefont{P.}~\bibnamefont{Kohlert}},
  \bibinfo{author}{\bibfnamefont{K.}~\bibnamefont{Kassner}}, \bibnamefont{and}
  \bibinfo{author}{\bibfnamefont{C.}~\bibnamefont{Misbah}},
  \bibinfo{journal}{Eur. Phys. J. B} \textbf{\bibinfo{volume}{35}},
  \bibinfo{pages}{493} (\bibinfo{year}{2003}).

\bibitem[{\citenamefont{Kassner et~al.}(2001)\citenamefont{Kassner, Misbah,
  M\"uller, Kappey, and Kohlert}}]{Kassner:01}
\bibinfo{author}{\bibfnamefont{K.}~\bibnamefont{Kassner}},
  \bibinfo{author}{\bibfnamefont{C.}~\bibnamefont{Misbah}},
  \bibinfo{author}{\bibfnamefont{J.}~\bibnamefont{M\"uller}},
  \bibinfo{author}{\bibfnamefont{J.}~\bibnamefont{Kappey}}, \bibnamefont{and}
  \bibinfo{author}{\bibfnamefont{P.}~\bibnamefont{Kohlert}},
  \bibinfo{journal}{Phys. Rev. E} \textbf{\bibinfo{volume}{63}},
  \bibinfo{pages}{036117} (\bibinfo{year}{2001}).

\bibitem[{\citenamefont{Torii and Balibar}(1992)}]{torii:92}
\bibinfo{author}{\bibfnamefont{R.}~\bibnamefont{Torii}} \bibnamefont{and}
  \bibinfo{author}{\bibfnamefont{S.}~\bibnamefont{Balibar}},
  \bibinfo{journal}{J. Low Temp. Phys.} \textbf{\bibinfo{volume}{89}},
  \bibinfo{pages}{391} (\bibinfo{year}{1992}).

\bibitem[{\citenamefont{Berr\'ehar et~al.}(1992)\citenamefont{Berr\'ehar,
  Caroli, Lapersonne-Meyer, and Schott}}]{Berrehar92}
\bibinfo{author}{\bibfnamefont{J.}~\bibnamefont{Berr\'ehar}},
  \bibinfo{author}{\bibfnamefont{C.}~\bibnamefont{Caroli}},
  \bibinfo{author}{\bibfnamefont{C.}~\bibnamefont{Lapersonne-Meyer}},
  \bibnamefont{and} \bibinfo{author}{\bibfnamefont{M.}~\bibnamefont{Schott}},
  \bibinfo{journal}{Phys. Rev. B} \textbf{\bibinfo{volume}{46}},
  \bibinfo{pages}{13487} (\bibinfo{year}{1992}).

\bibitem[{\citenamefont{Sekimoto and Kawasaki}(1989)}]{ken89}
\bibinfo{author}{\bibfnamefont{K.}~\bibnamefont{Sekimoto}} \bibnamefont{and}
  \bibinfo{author}{\bibfnamefont{K.}~\bibnamefont{Kawasaki}},
  \bibinfo{journal}{Physica A} \textbf{\bibinfo{volume}{154}},
  \bibinfo{pages}{384} (\bibinfo{year}{1989}).

\bibitem[{\citenamefont{Macosko}(1994)}]{Macosko}
\bibinfo{author}{\bibfnamefont{C.~W.} \bibnamefont{Macosko}},
  \emph{\bibinfo{title}{Rheology : principles, measurements, and applications}}
  (\bibinfo{publisher}{John Wiley \& Sons Inc}, \bibinfo{address}{New York},
  \bibinfo{year}{1994}).

\bibitem[{\citenamefont{He and Qiao}(2007)}]{HeEPL}
\bibinfo{author}{\bibfnamefont{L.~H.} \bibnamefont{He}} \bibnamefont{and}
  \bibinfo{author}{\bibfnamefont{L.}~\bibnamefont{Qiao}},
  \bibinfo{journal}{EPL} \textbf{\bibinfo{volume}{80}}, \bibinfo{pages}{14003}
  (\bibinfo{year}{2007}).

\bibitem[{\citenamefont{Tanaka et~al.}(1987)\citenamefont{Tanaka, Sun,
  Hirokawa, Katayama, Kucera, Hirose, and Amiya}}]{TanakaNat87}
\bibinfo{author}{\bibfnamefont{T.}~\bibnamefont{Tanaka}},
  \bibinfo{author}{\bibfnamefont{S.~T.} \bibnamefont{Sun}},
  \bibinfo{author}{\bibfnamefont{Y.}~\bibnamefont{Hirokawa}},
  \bibinfo{author}{\bibfnamefont{S.}~\bibnamefont{Katayama}},
  \bibinfo{author}{\bibfnamefont{J.}~\bibnamefont{Kucera}},
  \bibinfo{author}{\bibfnamefont{Y.}~\bibnamefont{Hirose}}, \bibnamefont{and}
  \bibinfo{author}{\bibfnamefont{T.}~\bibnamefont{Amiya}},
  \bibinfo{journal}{Nature} \textbf{\bibinfo{volume}{325}},
  \bibinfo{pages}{796} (\bibinfo{year}{1987}).

\bibitem[{\citenamefont{Oron et~al.}(1997)\citenamefont{Oron, Davis, and
  Bankoff}}]{Bankoff:1997}
\bibinfo{author}{\bibfnamefont{A.}~\bibnamefont{Oron}},
  \bibinfo{author}{\bibfnamefont{S.~H.} \bibnamefont{Davis}}, \bibnamefont{and}
  \bibinfo{author}{\bibfnamefont{S.~G.} \bibnamefont{Bankoff}},
  \bibinfo{journal}{Rev. Mod. Phys.} \textbf{\bibinfo{volume}{69}},
  \bibinfo{pages}{931} (\bibinfo{year}{1997}).

\bibitem[{\citenamefont{Spencer et~al.}(1991)\citenamefont{Spencer, Voorhees,
  and Davis}}]{Spencer91}
\bibinfo{author}{\bibfnamefont{B.~J.} \bibnamefont{Spencer}},
  \bibinfo{author}{\bibfnamefont{P.~W.} \bibnamefont{Voorhees}},
  \bibnamefont{and} \bibinfo{author}{\bibfnamefont{S.~H.} \bibnamefont{Davis}},
  \bibinfo{journal}{Phys. Rev. Lett.} \textbf{\bibinfo{volume}{67}},
  \bibinfo{pages}{3696} (\bibinfo{year}{1991}).

\bibitem[{\citenamefont{Mullins}(1959)}]{Mullins}
\bibinfo{author}{\bibfnamefont{W.~M.} \bibnamefont{Mullins}},
  \bibinfo{journal}{J. Appl. Phys.} \textbf{\bibinfo{volume}{30}},
  \bibinfo{pages}{77} (\bibinfo{year}{1959}).

\bibitem[{\citenamefont{M\"uller and Sa\'ul}(2004)}]{Saul}
\bibinfo{author}{\bibfnamefont{P.}~\bibnamefont{M\"uller}} \bibnamefont{and}
  \bibinfo{author}{\bibfnamefont{A.}~\bibnamefont{Sa\'ul}},
  \bibinfo{journal}{Surf. Sci. Rep.} \textbf{\bibinfo{volume}{54}},
  \bibinfo{pages}{157} (\bibinfo{year}{2004}).

\bibitem[{\citenamefont{Yang}(2005)}]{Yang05}
\bibinfo{author}{\bibfnamefont{F.}~\bibnamefont{Yang}}, \bibinfo{journal}{J.
  Phys. D: Appl. Phys.} \textbf{\bibinfo{volume}{38}}, \bibinfo{pages}{3938}
  (\bibinfo{year}{2005}).

\bibitem[{\citenamefont{Song and Yang}(2006)}]{Yang06}
\bibinfo{author}{\bibfnamefont{W.}~\bibnamefont{Song}} \bibnamefont{and}
  \bibinfo{author}{\bibfnamefont{F.}~\bibnamefont{Yang}}, \bibinfo{journal}{J.
  Phys. D: Appl. Phys.} \textbf{\bibinfo{volume}{39}}, \bibinfo{pages}{4634}
  (\bibinfo{year}{2006}).

\bibitem[{\citenamefont{Onuki}(1995)}]{OnukiPA1995}
\bibinfo{author}{\bibfnamefont{A.}~\bibnamefont{Onuki}},
  \bibinfo{journal}{Physica A} \textbf{\bibinfo{volume}{217}},
  \bibinfo{pages}{38} (\bibinfo{year}{1995}).

\bibitem[{\citenamefont{Pan et~al.}(2009)\citenamefont{Pan, Huang, Yu, and
  Feng}}]{Pan09}
\bibinfo{author}{\bibfnamefont{X.~H.} \bibnamefont{Pan}},
  \bibinfo{author}{\bibfnamefont{S.~Q.} \bibnamefont{Huang}},
  \bibinfo{author}{\bibfnamefont{S.~W.} \bibnamefont{Yu}}, \bibnamefont{and}
  \bibinfo{author}{\bibfnamefont{X.~Q.} \bibnamefont{Feng}},
  \bibinfo{journal}{J. Phys. D: Appl. Phys.} \textbf{\bibinfo{volume}{42}},
  \bibinfo{pages}{055302} (\bibinfo{year}{2009}).

\bibitem[{\citenamefont{de~Gennes}(2002)}]{deGennes02}
\bibinfo{author}{\bibfnamefont{P.~G.}~\bibnamefont{de~Gennes}},
  \bibinfo{journal}{Eur. Phys. J. E} \textbf{\bibinfo{volume}{7}},
  \bibinfo{pages}{31} (\bibinfo{year}{2002}).

\end{thebibliography}

\end{document}